\documentclass[12pt,letter]{article}
\pdfoutput=1
\usepackage{graphicx, epsfig, color,cite}
\usepackage{amsmath}
\usepackage{amssymb}
\usepackage{caption,subcaption,graphicx}
\usepackage[hidelinks]{hyperref}
\def\eslt{\not\!\!\!{E_T}}
\def\to{\rightarrow}

\def\bi{\begin{itemize}}
\def\ei{\end{itemize}}
\def\te{\tilde e}

\def\tchi{\tilde\chi}

\def\tu{\tilde u}

\def\tb{\tilde b}

\def\tst{\tilde t}
\def\ttau{\tilde \tau}

\def\tg{\tilde g}
\def\tnu{\tilde\nu}

\def\tw{\widetilde\chi^{\pm}}
\def\twns{\widetilde\chi}
\def\tz{\widetilde\chi^0}
\def\alt{\lesssim}
\def\agt{\gtrsim}
\def\be{\begin{equation}}  
\def\ee{\end{equation}}  
\def\bea{\begin{eqnarray}}  
\def\eea{\end{eqnarray}}  
\def\tbar{\bar{t}}

\begin{document}
\begin{titlepage}
\begin{flushright}
OU-HEP-210330
\end{flushright}

\vspace{0.5cm}
\begin{center}
{\Large \bf New angular (and other) cuts \\
to improve the higgsino signal at the LHC \\
}

\vspace{1.2cm} \renewcommand{\thefootnote}{\fnsymbol{footnote}}
{\large Howard Baer$^1$\footnote[1]{Email: baer@ou.edu },
Vernon Barger$^2$\footnote[2]{Email: barger@pheno.wisc.edu},
Dibyashree Sengupta$^3$\footnote[4]{Email: dsengupta@phys.ntu.edu.tw} 
and Xerxes Tata$^4$\footnote[5]{Email: tata@phys.hawaii.edu}
}\\ 
\vspace{1.2cm} \renewcommand{\thefootnote}{\arabic{footnote}}
{\it 
$^1$Homer L. Dodge Department of Physics and Astronomy,\\
University of Oklahoma, Norman, OK 73019, USA \\[3pt]
}
{\it 
$^2$Department of Physics,
University of Wisconsin, Madison, WI 53706 USA \\[3pt]
}
{\it 
$^3$Department of Physics,
National Taiwan University, Taipei, Taiwan 10617, R.O.C. \\[3pt]
}
{\it 
$^4$Department of Physics and Astronomy,
University of Hawaii, Honolulu, HI, USA\\[3pt]
}

\end{center}

\vspace{0.5cm}
\begin{abstract}
\noindent
Motivated by the fact that naturalness arguments strongly suggest that
the SUSY-preserving higgsino mass parameter $\mu$ cannot be too far
above the weak scale, we re-examine higgsino pair production in association
with a hard QCD jet at the HL-LHC. We focus on $\ell^+\ell^-+\eslt+j$
events from the production and subsequent decay,
$\tchi_2^0\to\tchi_1^0\ell^+\ell^-$, of the heavier neutral higgsino.  The
novel feature of our analysis is that we suggest angular cuts to
reduce the important background from $Z(\to \tau\tau)+j$ events more
efficiently than the $m_{\tau\tau}^2<0$ cut that has been used by the
ATLAS and CMS collaborations. Other cuts, needed to reduce backgrounds
from $t\bar{t}$, $WWj$ and $W/Z+\ell\bar{\ell}$ production, are also delineated.
We plot out the reach of LHC14 for 300 and 3000~fb$^{-1}$ and also show
distributions that serve to characterize the higgsino signal, noting
that higgsinos may well be the only superpartners accessible at LHC14 in
a well-motivated class of natural SUSY models.

\end{abstract}
\end{titlepage}

\section{Introduction}
\label{sec:intro}

\subsection{Motivation}

The discovery of a very Standard Model (SM)-like Higgs boson with mass
$m_h=125.10\pm 0.14$ GeV\cite{atlas_h,cms_h} at the CERN Large Hadron
Collider (LHC) is a great triumph. However, it also exacerbated a
long-known puzzle: what stabilizes the mass of a fundamental scalar
particle when quantum corrections should drive its mass far beyond its
measured value?\cite{susskind,veltman} The simplest and perhaps the most
elegant answer is that the weak scale effective field theory (EFT)
exhibits softly broken supersymmetry, and so has no quadratic
sensitivity to high scale physics \cite{witten_kaul}. The electroweak
scale is stabilized as long as soft supersymmetry breaking terms (at
least those involving sizeable couplings to the Higgs sector) are not
much larger than the TeV scale. The corresponding superpartners are then
expected to have masses around the weak scale \cite{wss}.
%
Up to now LHC superparticle searches\cite{canepa} have turned up
negative, resulting in lower  mass limits on the gluino of $m_{\tg}\agt 2.2$
TeV\cite{lhc_gl} and on the lightest top-squark $m_{\tst_1}\agt 1.1$
TeV\cite{lhc_t1}: these bounds are obtained within simplified models,
assuming that 1. the sparticle spectrum is not compressed, 2. $R$-parity is conserved and 3. gluinos and top-squarks dominantly decay to third
generation quarks/squarks (as expected in the scenarios considered
here\cite{Baer:1990sc}). Such strong
limits are well beyond early expectations for sparticle masses from
naturalness wherein $m_{\tg},\ m_{\tst_1}\alt 0.4$ TeV was expected
(assuming 3\% finetuning)\cite{eenz,bg,dg,ac}.\footnote{Naturalness
  bounds on gluino, top squark and other sparticle masses were
  historically derived using the Barbieri-Giudice (BG) measure\cite{eenz,bg}
  $\Delta_{EENZ,BG}$ by expressing $m_Z^2$ in terms of
  weak scale soft parameters $m_{H_u}^2$ and then expanding $m_{H_u}^2$
  in terms of high (GUT) scale parameters of the mSUGRA/CMSSM model
  using approximate semi-analytic solutions to the Minimal Supersymmetric
  Standard Model renormalization group equations. For further
discussion, see {\it e.g.} Ref. \cite{dew,mt,seige}.}
This disparity between
theoretical expectations and experimental reality has caused strong
doubts to be raised on the validity of the weak scale SUSY (WSS)
hypothesis\cite{dine}. While there is no question that supersymmetry
elegantly resolves the big hierarchy issue, 
the question often raised is: does WSS now suffer from a
Little Hierarchy Problem (LHP), wherein a putative mass
gap has opened up between the weak scale and the soft SUSY breaking
scale?

The LHP seemingly depends on how naturalness is measured in WSS.  The
original log-derivative measure \cite{eenz,bg} $\Delta_{BG}=max_i
|\partial\log m_Z^2/\partial\log p_i |$ (wherein the $p_i$ constitute
the various independent free parameters of the low energy effective
field theory in question), obviously depends on one's choice for these
parameters $p_i$.  In Ref's \cite{eenz,bg,dg,ac}, the EFT was chosen to
be constrained supersymmetric standard models (CMSSM or NUHM2) valid
up to energy scale $Q=m_{GUT}$ and the free parameters were taken to be
various GUT scale soft SUSY breaking terms such as common scalar mass
$m_0$, common gaugino mass $m_{1/2}$, common trilinear $A_0$ etc. The
various independent soft terms are introduced to {\it parametrize} our
ignorance of how SUSY breaking is felt by the superpartners of SM
particles.  However, if the CMSSM is derived from a more ultra-violet
complete theory ({\it e.g} string theory), then typically the EFT free
parameters are {\it determined} in terms of more fundamental parameters
such as the gravitino mass $m_{3/2}$ (in the case of
gravity-mediation). With a reduction in independent soft parameters,
parameters originally taken to be independent become correlated, and the
numerical fine-tuning value can change abruptly, even for exactly the
same numerical inputs\cite{dew,mt,seige}. Ignoring such correlations can
lead to an over-estimate of the fine-tuning by as much as two orders of
magnitude \cite{mt} and, perhaps, lead us to discard perfectly viable
models for the wrong reason.  An alternative measure, $\Delta_{HS}\sim \delta
m_{H_u}^2/m_h^2\sim \frac{3f_t^2}{16\pi^2}m_{\tst}^2\log (\Lambda^2
/m_{\tst}^2)$ (which favors top-squarks $m_{\tst_1}\alt 500$ GeV), turns
out to be greatly oversimplified in that it singles out one top-squark
loop contribution, again ignoring the possibility of underlying
cancellations in models with correlated parameters\cite{dew,mt,seige}.
%

A more conservative, parameter-independent measure $\Delta_{EW}$ was
proposed\cite{ltr,rns} which directly compares the magnitude of the weak
scale $m_Z^2$ to weak scale contributions from the SUSY Lagrangian:
\be
\frac{m_Z^2}{2}=\frac{m_{H_d}^2+\Sigma_d^d-(m_{H_u}^2+\Sigma_u^u)\tan^2\beta}{\tan^2\beta  -1}-\mu^2\simeq -m_{H_u}^2-\mu^2 -\Sigma_u^u(\tst_{1,2}) 
\label{eq:mzsq}
\ee
where $\Delta_{\rm EW}=max| largest\ RHS\ contribution|/(m_Z^2/2)$.
An upper limit on $\Delta_{\rm EW}$ (which we take to be $\Delta_{\rm
  EW} < 30$) then implies that the weak scale values of
$\sqrt{|m_{H_u}^2|}$ and $|\mu|$ should be $\alt 100-350$ GeV. This
means that the soft term $m_{H_u}^2$ is driven barely negative during
radiative EWSB (radiatively-driven natural SUSY, or RNS)\cite{ltr,rns}.
The SUSY-preserving $\mu$ term, which feeds mass to
$W,\ Z,\ h$ and {\it higgsinos}, is also in the $100-350$ GeV
range. Meanwhile, top-squark (and other sparticle) contributions to the
weak scale are loop suppressed and can lie in the $m_{\tst_1}\sim 1-3$
TeV range at little cost to naturalness\cite{upper,jaimie}.  Gluinos,
which influence the value of $m_Z$ mainly by their influence on the top
squark mass, can be as heavy as 6~TeV, or more\cite{upper,jaimie}. 
Thus, a quite natural spectrum emerges under $\Delta_{EW}$ wherein 
higgsinos lie at the lowest mass rungs, while stops, gluinos and electroweak gauginos may
comfortably lie within the several TeV range. First/second generation
squarks/sleptons may well lie in the 10-40 TeV range\cite{rns}. We
mention that (modulo technical caveats) $\Delta_{\rm EW} \le \Delta_{\rm
  BG}$, and further, that $\Delta_{\rm BG}$ reduces to $\Delta_{\rm EW}$
when it is computed with appropriate correlations between
high scale parameters \cite{dew,mt,seige}.

Although not connected directly to the main theme of this paper, we note
that it has been suggested that the RNS SUSY spectra are actually {\it to
  be expected} from considerations of the landscape of string theory
vacua, which also provides an understanding of the magnitude of the
cosmological constant $\Lambda_{cc}$\cite{weinberg,bp}.
Douglas\cite{doug}, Susskind\cite{suss} and Arkani-Hamed {\it et
  al.}\cite{adk} argue that large soft terms should be statistically
favored in the landscape by a power-law $f_{SUSY}(m_{soft})\sim
m_{soft}^{2n_F+n_D-1}$ where $n_F$ is the number of $F$-breaking fields
and $n_D$ is the number of $D$-breaking fields contributing to the
overall SUSY breaking scale. Thus, even for the textbook case of SUSY
breaking via a single $F$-term ($n_F=1,\ n_D=0$), there is already a
linear draw to large soft terms.  The landscape statistical draw to
large soft terms must be balanced by an anthropic requirement that EW
symmetry is properly broken (no charge-or-color (CCB) breaking minima in
the scalar potential and that EW symmetry is actually
broken)\cite{land0}.  Furthermore, if the value of $\mu$ is determined
by whatever solution to the SUSY $\mu$ problem is invoked\cite{mu}, then
$\mu$ is no longer available for finetuning and the {\it pocket
  universe} value of the weak scale $m_{weak}^{PU}$ should be within a
factor of a few of our universe's weak scale $m_{weak}^{OU}\simeq
m_{W,Z,h}\sim 100$ GeV. In pocket universes where $m_{weak}^{PU}$ is
larger than 4-5 times its observed value (remarkably, this corresponds to
$\Delta_{\rm EW}\alt 30$), Agrawal {\it et al.}\cite{agrawal} have shown
that nuclear physics goes awry, and atoms as we know them would not
form. Thus, one expects large (but not too large) soft SUSY breaking
terms, and consequently large sparticle masses (save higgsinos, which
gain mass differently).  Detailed calculations of Higgs and sparticle
masses find $m_h$ pulled to a statistical peak around $m_h\sim 125$ GeV
whilst sparticles other than higgsinos are pulled (well-) beyond current
LHC reach\cite{land0,land1,land2,land3}.

We stress that the top-down view of electroweak naturalness 
is mentioned only
by way of motivation and is in no way essential for the
phenomenological analysis of the higgsino signal studied in this
paper. The reader who does not subscribe to stringy naturalness can
simply ignore the previous paragraph. For that matter, even the
bottom-up naturalness considerations that led us  to focus on light
higgsinos do not play any essential role for the phenomenological
analysis that is suggested below. In other words, the reader not
interested in any naturalness considerations can
simply view the remainder of this paper as an improved analysis of 
how light higgsinos can be searched for at the high luminosity LHC. 

In our view, naturalness considerations make it very plausible that the best
hope for SUSY discovery at LHC is not via gluino or top-squark pair
production, but rather via light higgsino pair production: $pp\to
\twns_1^+\twns_1^-,\ \tz_1\tz_2,\ \tw_1\tz_2$.  While the total LHC
higgsino pair production cross section is substantial in the mass range
$\mu\sim 100-350$ GeV\cite{rns@lhc}, the problem is that very little
visible energy is released in higgsino decay $\tw_1\to
f\bar{f}^\prime\tz_1$ and $\tz_2\to f\bar{f}\tz_1$ (where $f$ stands for
SM fermions, for the most part $e$ and $\mu$ for the signals we study in
this paper) since most of the decay
energy ends up in the LSP rest mass $m_{\tz_1}$\cite{bbh}, unless binos
and winos are also fortituously light. Requiring that the higgsinos recoil
against hard initial state QCD radiation, not only provides an event
trigger but also boosts the higgsino decay products to measureable
energy values\cite{kribs,bmt,c_han}.  Indeed, much work has already
examined these reactions, and in fact limits have already been placed on
such signatures by the ATLAS\cite{atlas1,atlas2} and CMS\cite{cms1,cms2}
collaborations.

\subsection{Summary of some previous work and plan for this paper}

Here, we briefly summarize several previous studies on higgsino pair 
production and outline how the present work examines new 
territory.\footnote{There is a very substantial literature on {\it gaugino}
pair production signals at hadron colliders which we will not review here.
For a recent review on electroweakino searches at LHC, 
see \cite{Canepa:2020ntc}.}

\bi
\item In Ref. \cite{bbh}, higgsino pair production at LHC in the low
$\mu$ scenario was first examined. In that work, the reaction $pp\to\tz_1\tz_2$ 
with $\tz_2\to \ell^+\ell^-\tz_1$ was explored without requiring hard 
initial state radiation (ISR).
Instead, a soft dimuon trigger was advocated. With such a trigger, then signal 
and BG rates were found to be comparable and the search for collimated
opposite-sign/same-flavor (OS/SF) dileptons plus MET was advocated
where the signal would exhibit a characteristic bump in dilepton invariant
mass with $m(\ell^+\ell^- )<m_{\tz_2}-m_{\tz_1}$.

\item In Ref. \cite{kribs}, Han, Kribs, Martin and Menon examined the
  reaction $pp\to\tz_1\tz_2 j$, where the higgsinos recoiled against a
  hard QCD radiation.  A hard cut $m_{\tau\tau}^{\rm HKMM}>150$ GeV was
  used to reduce $Z\to\tau^+\tau^- j$ background. The bump in
  $m(\ell^+\ell^- )< m_{\tz_2}-m_{\tz_1}$ was displayed above SM BGs for
  several signal benchmark models.

\item In Ref. \cite{bmt}, an improved $m_{\tau\tau}^2$ variable was
  defined, with a crucial $m_{\tau\tau}^2<0$ cut used to reject
  $\tau\bar{\tau}j$ events compared to signal. A very conservative
  $b$-jet tag efficiency of 60\% resulted in a dominant $t\bar{t}$
  background. The current ATLAS $b$-tag efficiency is given at 85\% so
  that requiring no $b$-jets in BG events substantially reduces
  $t\bar{t}$ BG. Reach contours were plotted vs. $\mu$ for several
  values of $m_{1/2}$ assuming integrated luminosities up to 1000
  fb$^{-1}$ in this pre-HL-LHC paper. The reach plot was extended to
  3000 fb$^{-1}$ in Ref. \cite{Baer:2016usl}.

\item Ref. \cite{c_han} focused on SUSY models with $\Delta m^0\equiv
  m_{\tz_2}-m_{\tz_1}\alt 5$ GeV and the well-collimated dimuon pair was
  regarded as a single object $\mu_{col}$.  Hard $\eslt>250$ GeV and
  $p_T(jet)>250$ GeV cuts were applied along with transverse mass
  $m_T(\mu_{com},\eslt )<50$ GeV and $\eslt / p_T(\mu_{col})>20$.
  Significance $S/\sqrt{BG}$ for three examined BM points were found to
  range from $1.85-2.9\sigma$ for assumed integrated luminosity of 3000
  fb$^{-1}$.

\item The CMS collaboration examined the soft dilepton+jet+$\eslt$
  signature in Ref. \cite{cms1} using 35.9 fb$^{-1}$ of data at
  $\sqrt{s}=13$ TeV.  They were able to exclude values of $m_{\tz_2}$ up
  to about 167 GeV for $\Delta m^0\sim 15$ GeV although the limit drops
  off as $\Delta m^0$ falls off below or above this central value. A
  follow-up paper using 139 fb$^{-1}$ of data at 13 TeV extended these
  limits up to $\mu\sim 200$ GeV\cite{cms2}.

\item ATLAS examined the soft dilepton+jet+$\eslt$ signature in Ref's
  \cite{atlas1} using 36.1 fb$^{-1}$ of data at $\sqrt{s}=13$ TeV where
  they reported the utility of an $\eslt/H_T(\ell )\agt 5$ cut.  They
  updated their search to 139 fb$^{-1}$ in \cite{atlas2}.  In the latter
  paper, values of $m_{\tz_2}\alt 200$ GeV were excluded for $\Delta
  m^0\sim 10$ GeV with a rapid drop-off below and above this value. Some
  signal excess was noted for low $m(\ell^+\ell^-)\sim 4-12$ GeV for
  their signal region SR-E-med plot.

\item In Ref. \cite{Baer:2020sgm}, theoretical aspects of the higgsino
  discovery plane $m_{\tz_2}$ vs. $\Delta m^0$ were explored.  It was
  shown that the string landscape prefers the smaller mass gap region
  $\Delta m^0\sim 4-12$ GeV with $m_{\tz_2}\sim 100-350$ GeV.  In
  contrast, the LHC limit on the gluino mass constrains natural models
  with gaugino mass unification to have $\Delta m^0 \sim 10-25$~GeV. 

\ei

Our goal in the present paper is to re-examine the promising soft OS/SF
dilepton plus jets plus $\eslt$ signal in light of its emerging
strategic importance for natural SUSY discovery in the HL-LHC era.  We
provide a detailed characterization of both expected signal and dominant
SM backgrounds by displaying a wide variety of distributions of various
kinematic variables.  We also suggest new angular cuts that are much
more efficient than the currently used $m_{\tau\tau}^2<0$ cut in
suppressing the important SM background from $Z (\rightarrow
\tau\bar{\tau})+ jet$ production, thus aiding in the signal search at
the HL-LHC.


\section{Natural SUSY benchmark points}
\label{sec:BMpoints}

In this section, we delineate three SUSY benchmark points (BM) that
are used throughout the paper in order to compare signal strength
against SM background rates. We use the computer code Isajet
7.88\cite{isajet} to generate all sparticle mass spectra. The ensuing
SUSY Les Houches Accord (SLHA) files are input to
Madgraph\cite{madgraph}/Pythia\cite{pythia}/Delphes\cite{delphes}
for event generation.
We select points for
varying higgsino masses, and equally importantly, with different
neutralino-LSP mass gaps $\sim 4-16$~GeV. The three BM points are listed
in Table \ref{tab:bm}.

Our first BM point is listed as BM1 in Table \ref{tab:bm}.  It is
generated from the two-extra-parameter non-universal Higgs model (NUHM2)
with parameters $m_0,\ m_{1/2},\ A_0,\ \tan\beta,\ \mu,\ m_A$
$=5000\ {\rm GeV},\ 1001\ {\rm GeV},\ -8000\ {\rm GeV},\ 10,\ 150\ {\rm
  GeV},\ 2000\ {\rm GeV}$. It has $m_{\tg}\sim 2.4$ TeV and
$m_{\tst_1}\sim 1.6$ TeV so is LHC allowed via gluino and top squark
searches. With a relatively small
value $\mu=150$ GeV and a sizeable neutralino mass gap
$\Delta m^0 \sim 12$ GeV, it is just within the 95\% CL region
now excluded by ATLAS\cite{atlas2} and CMS\cite{cms2} soft dilepton
searches.  It is natural in that $\Delta_{EW}\sim 14$.

Our second BM point (denoted BM2) is also from NUHM2 model. 
It has $\mu= 300$ GeV  with a mass gap 
$m_{\tz_2}-m_{\tz_1}\sim 16$ GeV so is well beyond current 
ATLAS/CMS search limits for soft dileptons+jets+$\eslt$. It has 
$\Delta_{EW}\sim 22$.

Our third point, listed as BM3 (GMM$^\prime$), comes from natural
generalized mirage mediation model\cite{ngmm} where $\mu$ is used as an
input (GMM$^\prime$). This model combines moduli/gravity-mediation with
anomaly mediated SUSY breaking (AMSB) via a mixing factor $\alpha$,
where $\alpha\to 0$ corresponds to pure AMSB and $\alpha\to\infty$
corresponds to pure gravity-mediation.  It uses the gravitino mass
$m_{3/2}=75$ TeV as input along with continuous factors $c_m$, $c_{m3}$
and $a_3$ related to the generation 1,2 scalar masses, generation 3
scalar masses and $A$ parameters, respectively\cite{ngmm}.  We take $\mu
=200$ GeV. Since the gaugino masses unify at the intermediate mirage
unification scale $\mu_{mir}\sim 5.3\times 10^7$ GeV, then for a given
gluino mass, the wino and bino masses will be much heavier as compared
to models with unified gaugino masses such as NUHM2.  This means the
corresponding neutralino mass gap $m_{\tz_2}-m_{\tz_1}\sim 4.3$ GeV so
that the $\tz_2$ decay products will be very soft, making its search a
challenge even though higgsinos are not particularly heavy. The model
yields $\Delta_{EW}=26$.

Although outside of the main theme of the paper, we also list values for
some low energy and dark-matter-related observables towards the bottom
of Table \ref{tab:bm}.\footnote{The relic abundance of thermally-produced
  higgsino-like WIMPs listed in Table~\ref{tab:bm} are a factor of
  17, 5 and 13 below the measured dark matter abundance $\Omega_{DM}h^2=0.12$
  for each of benchmark points BM1, BM2 and BM3, respectively.
  The remaining abundance might be made of a second dark matter particle
  such as axions. With such a reduced abundance of higgsino-like WIMPs,
  then higgsino-like WIMPs are still allowed DM candidates even in the
  face of constraints from indirect dark matter detection
  experiments\cite{Baer:2018rhs}.}  

\begin{table}\centering
\begin{tabular}{lccc}
\hline
parameter & $BM1$ & $BM2$ & $BM3\ (GMM^\prime)$\\
\hline
$m_0$        & 5000 & 5000 & $\textendash$ \\
$m_{1/2}$     & 1001 & 1000 & $\textendash$ \\
$A_0$        & -8000 & -8000 & $\textendash$ \\
$\tan\beta$  & 10 & 10 & 10 \\
$\mu$          & 150   & 300 & 200 \\
$m_A$          & 2000  & 2000 & 2000 \\
\hline
$m_{3/2}$     & $\textendash$ & $\textendash$ & 75000  \\
$\alpha$     & $\textendash$ & $\textendash$ & 4 \\
$c_m$        & $\textendash$ & $\textendash$ & 6.9 \\
$c_{m3}$      & $\textendash$ & $\textendash$ & 6.9 \\
$a_3$        & $\textendash$ & $\textendash$ & 5.1 \\
\hline
$m_{\tg}$   & 2425.4  & 2422.6 & 2837.3 \\
$m_{\tu_L}$ & 5295.9 & 5295.1 & 5244.6 \\
$m_{\tu_R}$ & 5427.8 & 5426.5 & 5378.0 \\
$m_{\te_R}$ & 4823.7  & 4824.5 & 4813.2 \\
$m_{\tst_1}$ & 1571.7 & 1578.4 & 1386.9  \\
$m_{\tst_2}$ & 3772.0 & 3773.0 & 3716.7 \\
$m_{\tb_1}$ & 3806.7 & 3807.6 & 3757.8 \\
$m_{\tb_2}$ & 5161.2 & 5160.2 & 5107.7 \\
$m_{\ttau_1}$ & 4746.8 & 4747.5 & 4729.8 \\
$m_{\ttau_2}$ & 5088.6 & 5088.2 & 5075.7 \\
$m_{\tnu_{\tau}}$ & 5095.4 & 5095.0 & 5084.8 \\
$m_{\tw_2}$ & 857.1 & 857.6 & 1801.9 \\
$m_{\tw_1}$ & 156.6 & 311.6 & 211.1 \\
$m_{\tz_4}$ & 869.0 & 869.8 & 1809.3 \\ 
$m_{\tz_3}$ & 451.3 & 454.7 & 1554.4 \\ 
$m_{\tz_2}$ & 157.6 & 310.1 & 207.0 \\ 
$m_{\tz_1}$ & 145.4 & 293.7 & 202.7 \\ 
$m_h$       & 124.5 & 124.6 & 125.4 \\ 
\hline
$\Omega_{\tz_1}^{std}h^2$ & 0.007 & 0.023 & 0.009  \\
$BF(b\to s\gamma)\times 10^4$ & 3.1 & 3.1 & 3.1 \\
$BF(B_s\to \mu^+\mu^-)\times 10^9$ & 3.8 & 3.8 & 3.8\\
$\sigma^{SI}(\tz_1 p)$ (pb) & $0.23\times10^{-8}$ & $0.52\times10^{-8}$ & $0.30\times10^{-9}$ \\
$\sigma^{SD}(\tz_1 p)$ (pb) & $0.86\times10^{-4}$ & $0.49\times10^{-4}$ &  $0.54\times10^{-5}$  \\
$\langle\sigma v\rangle |_{v\to 0}$  (cm$^3$/sec)  & $0.3\times10^{-24}$ & $0.1\times10^{-24}$ &  
$0.2\times10^{-24}$ \\
$\Delta_{\rm EW}$ & 13.9 & 21.7 & 26.0 \\
\hline
\end{tabular}
\caption{Input parameters and masses in~GeV units for two NUHM2 model
  benchmark points (BM1 and BM2)  and one natural mirage mediation SUSY
  benchmark point (BM3 (GMM')),
  with $m_t=173.2$ GeV. The input parameters for the
  natural(generalized) mirage mediation model such as $\alpha$ and $c_m$
  have been calculated from $m_0^{MM}$ and $m_{1/2}^{MM}$ which are
  taken equal to the corresponding NUHM2 model values of $m_0$ and
  $m_{1/2}$, respectively.  The $c_m$ and $c_{m3}$ have been taken equal
  to each other so that masses of first/second and third generation
  sfermions are equal at the GUT scale so as to also match the NUHM2
  models in the second and third columns of the table.
}
\label{tab:bm}
\end{table}

\section{Calculational details}
\label{sec:calc}

\subsection{Event generation}
\label{ssec:evgen}

$pp$ collision events with $\sqrt{s}=14$~TeV were generated using
{\sc MadGraph}~2.5.5~\cite{madgraph} interfaced to {\sc PYTHIA}
v8~\cite{pythia} via the default MadGraph/PYTHIA interface with default
parameters for showering and hadronization.  Detector simulation is
performed by {\sc Delphes} using the default
Delphes~3.4.2~\cite{delphes} ``ATLAS'' parameter card.

We utilize the anti-$k_T$ jet algorithm~\cite{Cacciari:2008gp}
with $R = 0.6$ (the default value in the ATLAS Delphes card) rather than
the 
Delphes card default value, $R = 0.5$.  (Jet finding in
Delphes is implemented via {\sc FastJet}~\cite{Cacciari:2011ma}.)  We
consider only jets with transverse energy satisfying $E_T(jet) > 40$ GeV
and pseudorapidity satisfying $|\eta(jet)| < 3.0$ in our analysis.
We implement the default Delphes $b$-jet tagger and implement a 
$b$-tag efficiency of 85\%~\cite{ATLAS:2015dex}.
 
The lepton identification criteria that we adopt are modified from the
default version of Delphes. We identify leptons with $E_T> 5$~GeV and
within $|\eta (\ell )| < 2.5$. We label them as isolated leptons if the
sum of the transverse energy of all other objects (tracks, calorimeter
towers, etc.) within $\Delta R = 0.5$ of the lepton candidate is less
than $10\%$ of the lepton $E_T$.

\subsection{SM background processes}

Using Madgraph-Pythia-Delphes, we generate $10^5$ signal events for each of the
Table \ref{tab:bm} benchmark points. We also evaluated SM backgrounds from
\bi
\item $\tau\bar{\tau}j$ production,
\item $t\bar{t}$ production,
\item $WWj$ production,
\item $W\ell\bar{\ell}j$ production, and
\item $Z\ell\bar{\ell}j$ production, 
\ei 
generating $10^5$ events for
  each of the background processes except $\tau\bar{\tau}j$ and $t\bar{t}$ where we
  generate $10^6$ events and also force both the tops to decay into $e$,
  $\mu$ or $\tau$ leptons for the latter.  For the processes containing
  $\ell\bar{\ell}$ (here, $\ell=e,\mu$ or $\tau$) the lepton pair is
  produced via the decay of a virtual photon or a $Z$-boson.
  For the $\tau\bar{\tau}j$ background, we allow for all possible $\tau$
  decay modes and then pick out the soft same-flavor opposite sign
  dilepton pairs at the toy detector simulation (Delphes) level.
%
%

\section{Higgsino signal analysis and  SM backgrounds}
\label{sec:distributions}

For the SUSY signal from higgsinos, we generate events from the
reactions $pp\to\twns_1^\pm\tz_2$, $\tz_1\tz_2$ and $\twns_1^+\twns^-$
where $\tz_2\to \tz_1\ell^+\ell^-$. The visible decay products from
$\twns_1^\pm$ and $\tz_2$ decays are typically soft because of their
small mass difference with the LSP.

\subsection{Parton level cuts and $C1$ cuts} \label{subsec:c1cuts}

Our listing of the dilepton plus jet signal and various background cross
sections after a series of cuts detailed below is shown in Table
\ref{tab:xsec}.  The first entry  labeled $BC$ for
    {\it before cuts} actually has parton level cuts implemented (at the
    Madgraph level) since some of the subprocesses are otherwise
    divergent. Also, for the backgrounds with a hard QCD ISR (labeled as
    $j$ in row 1), we require $p_T(j)>80$ GeV to efficiently generate
    events with a hard jet. For the backgrounds including
    $\gamma^*,Z^*\to\ell\bar{\ell}$ ($\ell=e$ or $\mu$), we implement
    $m(\ell\bar{\ell})>1$ GeV to regularize the otherwise divergent
    photon propagator. We also require $p_T(\ell )>1$ GeV and $\Delta
    R(\ell\bar{\ell})>0.01$, again at the parton level.  The $W$
    daughters of top quarks in $t\bar{t}$ events are forced to decay
    leptonically (into $e$, $\mu$ or $\tau$), but not so the $W$-bosons
    in first entry of the $WWj$ column.  These parton events are then
    fed into PYTHIA and analysed using the DELPHES detector simulation.
    The leading order cross sections (in $fb$), for both the signal as
    well as for the background, are listed in row 2 and labelled as
    $BC$.  Here, we see the signal reactions lie in the 10-100 fb regime
    whilst SM backgrounds are dominated by $t\bar{t}$ and
    $\tau\bar{\tau}j$ production and are about 500 times larger than
    signal point BM1.

To select out signal events, we implement cut set {\bf C1}:
\bi
\item require two opposite sign, same flavour (OS/SF) isolated leptons
  with $p_T(\ell )>5$ GeV, $|\eta (\ell )|<2.5$,
\item require there be at least one jet in the event; {\it i.e.}, $n_j\ge
  1$ with $p_T(j_1)>100$ GeV for identified calorimeter jets,
\item require $\Delta R(\ell\bar{\ell})>0.05$ (for $\ell =e$ or $\mu$),
\item require $\eslt >100$ GeV, and
\item veto tagged $b$-jets, $n$($b$-jet)=0.
\ei

\begin{table}\centering
\begin{tabular}{lcccccccc}
\hline
cuts/process & $BM1$ & $BM2$ & $BM3 (GMM^\prime)$ & $\tau\bar{\tau}j$ & 
$t\bar{t}$ & $WWj$ & $W\ell\bar{\ell}j$ & $Z\ell\bar{\ell}j$ \\
\hline
$BC$        & 83.1 & 9.3 & 31.3 & 43800.0 & 41400 & 9860 & 1150.0 & 311 \\
$C1$        & 1.2 & 0.19 & 0.07 & 94.2 & 179 & 35.9 & 14.7 & 5.9 \\
$C1+m_{\tau\tau}^2<0$        & 0.92 & 0.13 & 0.043 & 23.1 & 75.6 & 12.8 & 7.7 
& 3.2 \\
$C1+angle$        & 0.69 & 0.12 & 0.04 & 2.2 & 130 & 22.1 & 11.0 & 4.9 \\
$C2$        & 0.29 & 0.049 & 0.019 & 0.13 & 0.99 & 0.49 & 0.18 & 0.14 \\
$C3$        & 0.25 & 0.033 & 0.017 & 0.13 & 0.29 & 0.39 & 0.15 & 0.07 \\
\hline
\end{tabular}
\caption{Cross sections (in $fb$) for signal benchmark points and the
  various SM backgrounds listed in the text after various cuts. The row
  labelled BC denotes parton level cross sections after  the requirement
  $p_T(j)> 80$~GeV, along with
minimal cuts
  implemented to regulate divergences, and also includes the leptonic
  branching fractions for decays of both the top quarks in the
  $t\bar{t}$ column. The remaining rows list the cross sections after a
  series of analysis cuts detailed in the text. }
\label{tab:xsec}
\end{table}

After {\bf C1} cuts, signal cross sections for higgsino events with
exactly two OS/SF isolated leptons plus at least one jet with $P_T>
100$~GeV and $\eslt> 100$~GeV, are at the $fb$ or below level while
corresponding SM backgrounds lie in the $5-200$~fb range.
Note that after each set of cuts, of the three BM points,
BM3 has the lowest surviving signal cross section as a consequence of its
tiniest $\Delta m^0$ mass gap which leads to very soft leptons from
$\tchi_2^0$ decay.

\subsection{$m_{\tau\tau}^2$ vs. new angular cuts}
\label{ssec:angle}

\subsubsection{$m_{\tau\tau}^2$ cut}
\begin{figure}[!htbp]
\begin{center}
\includegraphics[height=0.4\textheight]{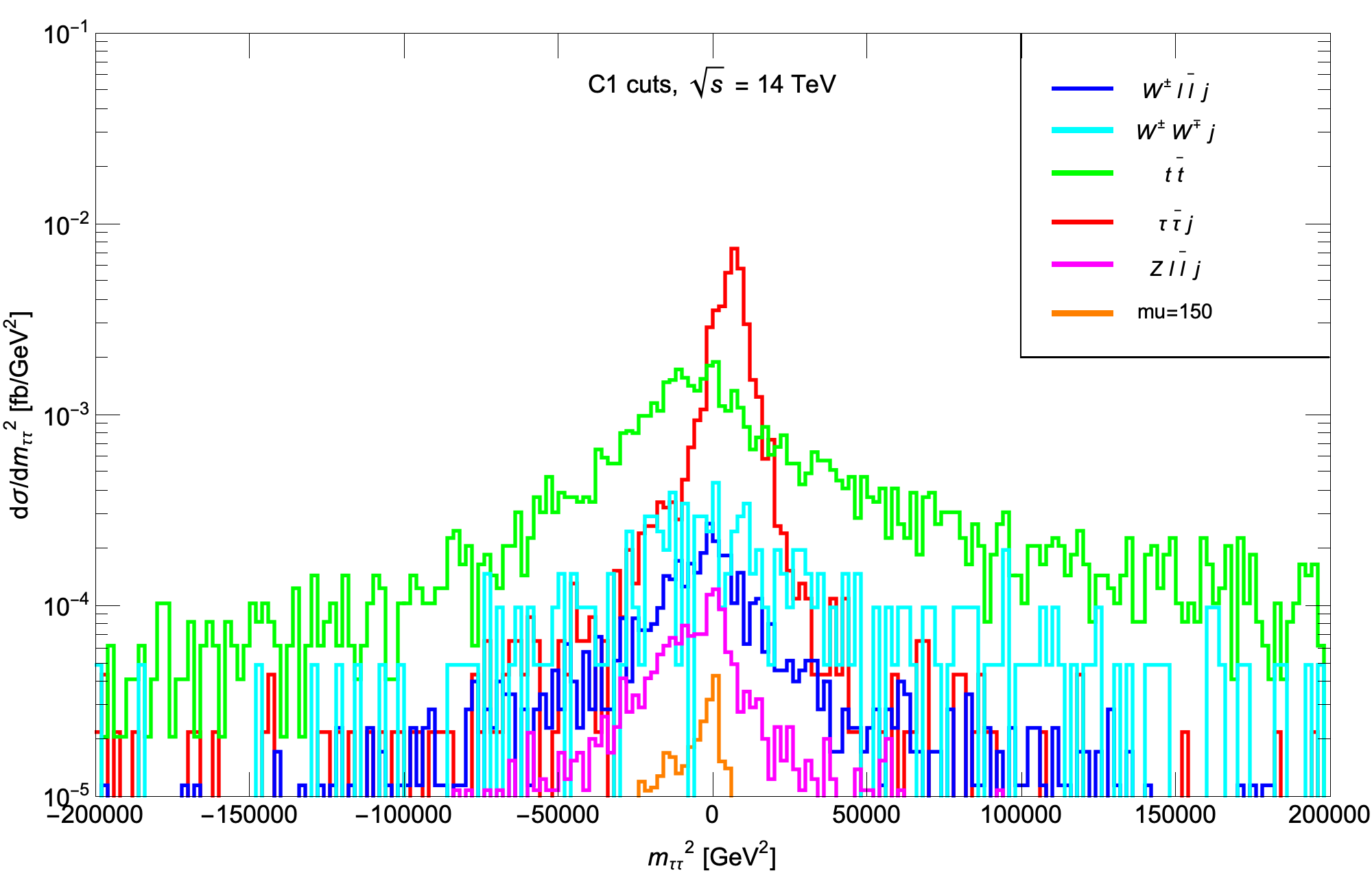}
\caption{Distribution in $m_{\tau\tau}^2$ for the three SUSY BM models
  with $\mu =150,\ 200$ and 300 GeV introduced in the text, along with
  SM backgrounds after $C1$ cuts augmented by $n_j=1$.
\label{fig:mtt}}
\end{center}
\end{figure}

We see from Table~\ref{tab:xsec} that $\tau\bar{\tau}j$ and $t\bar{t}$
processes constitute the largest backgrounds after C1 cuts. For the
most part, hard taus come from the decay of an on-shell high $p_T$ $Z$
boson recoiling against a hard QCD jet, and so are very relativistic.
In the approximation that the leptons and neutrinos from the decay of
each tau are all exactly collimated along the parent tau direction, we
can write the momentum carried off by the two neutrinos from the decay
$\tau_1\to \ell_1\bar{\nu}_{\ell_1}\nu_{\tau_1}$ of the first tau as
$\xi_1\vec{p}(\ell_1)$ and, similarly, as $\xi_2\vec{p}(\ell_2)$ for the
second tau.  Momentum conservation in the plane transverse to the beams
then requires that \be
-\sum_{jets}\vec{p}_T(j)=(1+\xi_1)\vec{p}_T(\ell_1
)+(1+\xi_2)\vec{p}_T(\ell_2 ) .
\label{eq:jetsum}
\ee 
These two equations can be solved for $\xi_1$ and $\xi_2$ given that
$\vec{p}_T(j)$ and $\vec{p}_T(\ell_{1,2})$ are all measured, and used to
evaluate the momenta of the individual taus.  This then allows us to
evaluate the invariant mass squared of the di-tau system which (within
the collinear approximation for tau decays) is given by,
\be
m_{\tau\tau}^2=(1+\xi_1)(1+\xi_2)m_{\ell\ell}^2 .
\ee
We show the distribution of $m_{\tau\tau}^2$ for both signal events as
well as for the various backgrounds in Fig. \ref{fig:mtt} after the cut
set {\bf C1} and further imposing $n_j=1$.\footnote{We make this
  additional requirement because, as we will see in
  Sec.~\ref{subsec:c2cuts}, limiting $n_j$ to be one helps to greatly
  reduce the $t\bar{t}$ background.}  As expected, this peaks sharply
around $m_Z^2$ for the $\tau\bar{\tau}j$ background (red histogram). In
contrast, for signal and other SM background events, where the isolated
lepton and $\vec{\eslt}$ directions are uncorrelated, the
$m_{\tau\tau}^2$ distributions are very broad and peak at even negative
values.  Thus, the $m_{\tau\tau}^2$ provides a very good discriminator
between $\tau\bar{\tau}j$ background and signal, and has, in fact, been used
in ATLAS \cite{atlas2} and CMS \cite{cms2} for their analyses.  We see,
however, that a rather extensive tail from the $\tau\bar{\tau}j$
background extends to negative values and arises due to tau pair
production from virtual photons, the breakdown of the collinear
approximation for asymmetric $Z$ decays and finally hadronic energy
mismeasurements which skew the direction of both $\vec{p}_T(j)$ and of
$\vec{\eslt}$.  Thus, in accord with Ref.~\cite{bmt}, we will require
$m_{\tau\tau}^2<0$ in the fourth row of Table \ref{tab:xsec} after ${\bf
  C1}$ cuts. We see that the ditau background is reduced by a factor 4
in contrast to the signal which is reduced by 25-40\%, depending on the
benchmark point.

Even after the $m_{\tau\tau}^2<0$ cut, substantial $\tau\bar{\tau}j$
background remains. We have checked that after additional cuts
(described in the next section) to reduce the $t\bar{t}$ background,
$\tau\bar{\tau}j$ production remains as the dominant irreducible
background.\footnote{We do not show these results for brevity.} This is
in sharp contrast to the analysis in Ref. \cite{bmt} where $t\bar{t}$
production remained as the dominant physics background even after the
$m_{\tau\tau}^2 < 0$ cut.  It is mainly the stronger $b$-jet veto
attained by ATLAS/CMS along with further cuts described below that leads
in the present case to $\tau\bar{\tau}j$ production as the dominant
background.  This motivated us to examine whether it is possible to
reduce the di-tau background more efficiently, without a huge loss of
signal. We turn to a discussion of this in Sec.\ref{subsubsec:angle}.

\subsubsection{New angle cuts} \label{subsubsec:angle}

\begin{figure}[!htbp]
\begin{center}
\includegraphics[height=0.4\textheight]{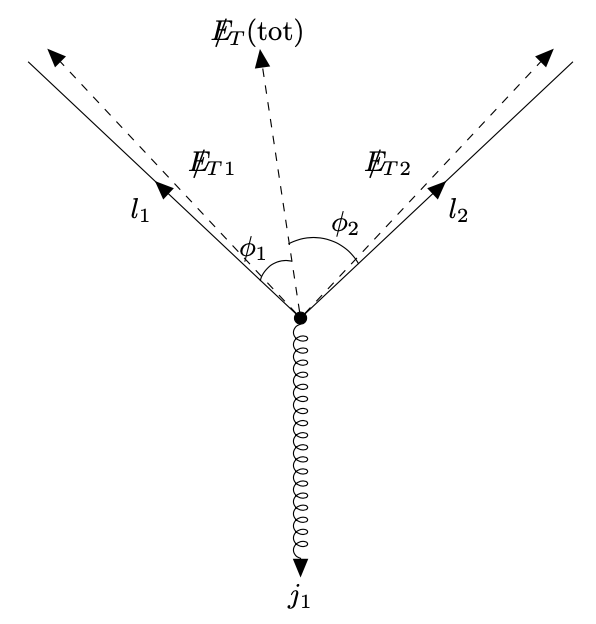}
\caption{Sketch of a ditau background event to the di-lepton plus jet plus
$\eslt$ signature in the transverse plane of the event. Here $\ell_1$ and
  $\eslt_{1}$ denote the transverse momentum of the lepton and of the
  vector sum of the neutrinos from the decay of the first tau, and
  likewise $\ell_2$ and $\eslt_2$. $\eslt$(tot) is the resultant $\eslt$
  in the event. Notice that because the taus are expected to be
  relativistic, $\ell_i$ and $\eslt_i$ vectors are nearly collimated
  along the direction of the $i^{th}$ tau ($i=1,2$). 
\label{fig:sketch}}
\end{center}
\end{figure}

In this subsection, we propose new angular cuts to replace the
$m_{\tau\tau}^2<0$ cut that we have just discussed. In the transverse
plane, the di-tau pair must recoil against the hard QCD radiation with
an opening angle between the taus significantly smaller than $\pi$. The
central idea, illustrated in Fig.~\ref{fig:sketch}, is that the $\eslt$
vector {\it must lie between the directions of the two taus} which (for
relativistic taus) are, of course, essentially the same as the {\it
  observable} directions of the charged lepton daughters of the taus. We
require the azimuthal angles $\phi_\ell$ and $\phi_{\bar{\ell}}$ for
each lepton to lie between $0$ and $2\pi$, and define
$\phi_{max}=max(\phi_\ell,\phi_{\bar{\ell}})$ and
$\phi_{min}=min(\phi_\ell,\phi_{\bar{\ell}})$. Then for $\vec{\eslt}$ to
lie in between the tau daughter lepton directions we must have,
\footnote{This works as long as $|\phi_{\ell}-\phi_{\bar{\ell}}|< \pi$. If
  $|\phi_{\ell}-\phi_{\bar{\ell}}|> \pi$, define $\phi_{\ell}^\prime
  =\phi_{\ell}+\pi$, $\phi_{\bar{\ell}}^\prime= \phi_{\bar{\ell}}+\pi$ and
  $\phi_{\hspace{1mm} \eslt}^\prime =\phi_{\hspace{1mm} \eslt}+\pi$, (all
  modulo $2\pi$) along with
  $\phi_{max}=max(\phi_\ell^\prime,\phi_{\bar{\ell}}^\prime)$, and likewise,
  $\phi_{min}=min(\phi_\ell^\prime,\phi_{\bar{\ell}}^\prime)$, and then require,
  $\phi_{min} < \phi_{\hspace{1mm} \eslt} < \phi_{max}$.}
%
%
$$\phi_{min} < \phi_{\hspace{1mm} \eslt}<\phi_{max}.$$ Notice that, by
definition, $\phi_{max}-\phi_{min} < \pi$, and for a boosted tau pair,
often significantly smaller than $\pi$.

\begin{figure}[!htbp]
\begin{center}
\includegraphics[height=0.4\textheight]{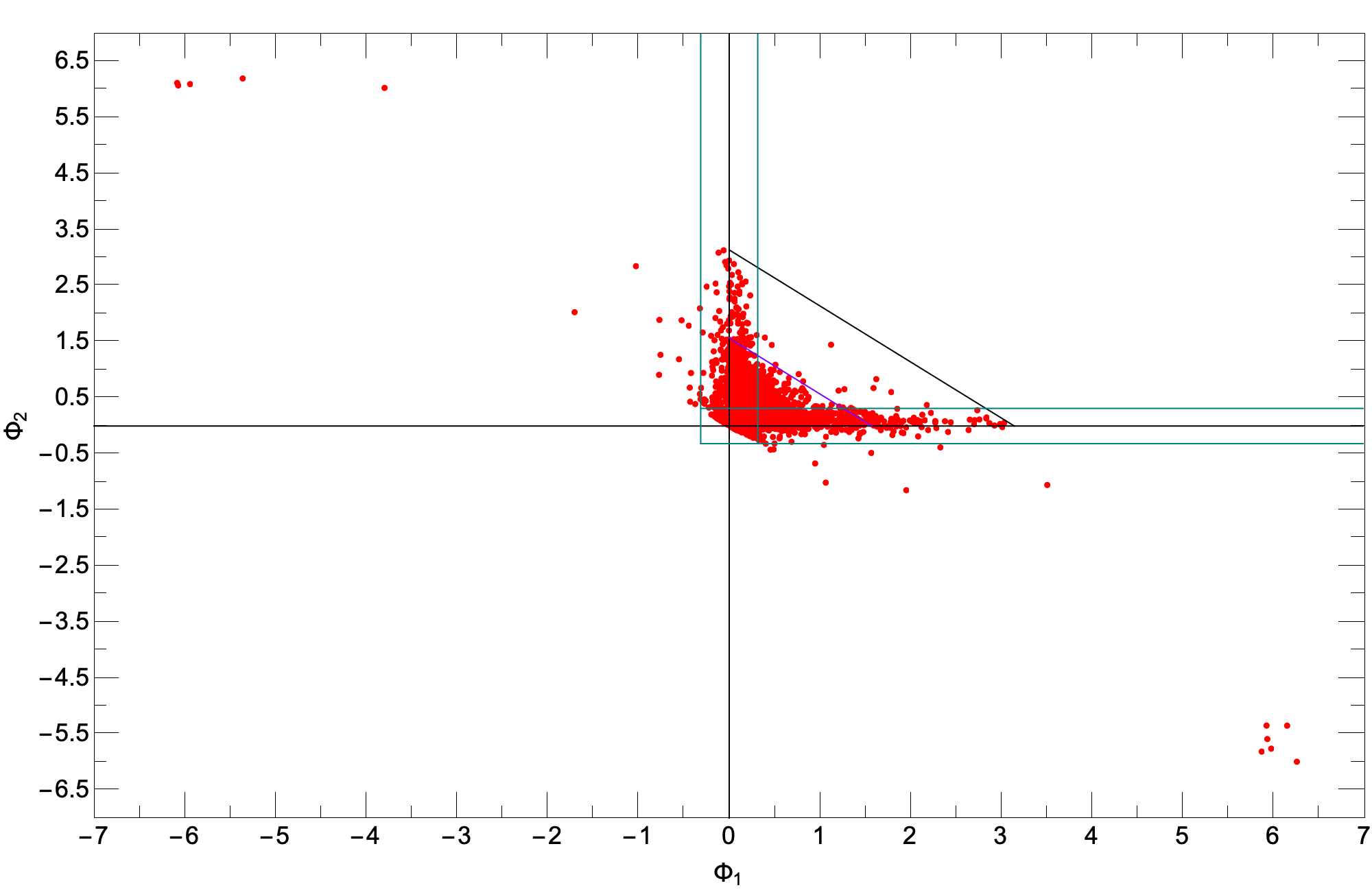}
\caption{Distribution in $\phi_1$ vs. $\phi_2$ plane for 
$\tau\bar{\tau}j$ background after $C1$ cuts, rquiring also that $n_j=1$.
\label{fig:phi1pihi2_tautauj}}
\end{center}
\end{figure}

\begin{figure}[!htbp]
\begin{center}
\includegraphics[height=0.4\textheight]{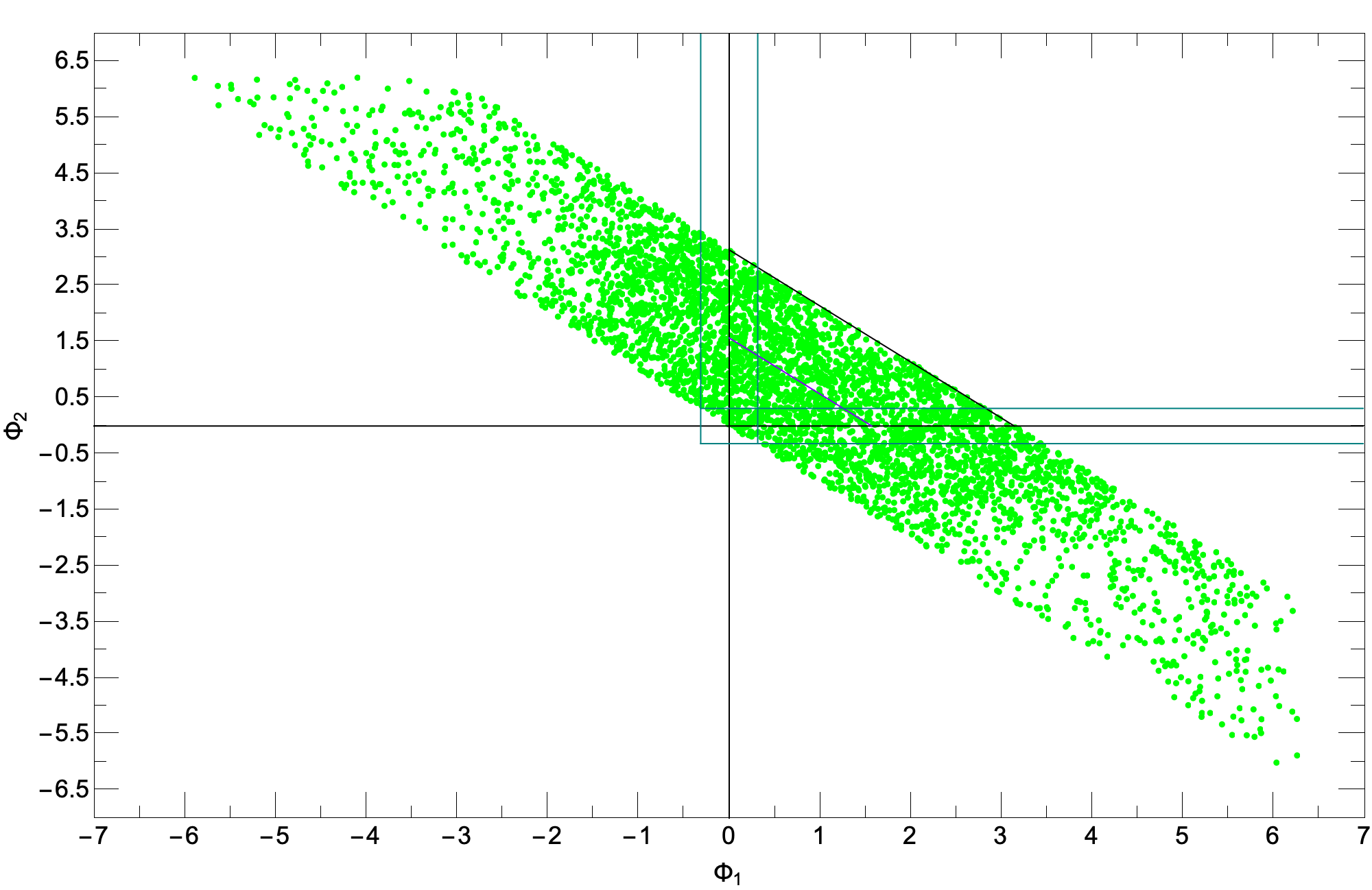}
\caption{Distribution in $\phi_1$ vs. $\phi_2$ plane for 
$t\bar{t}$ background after $C1$ cuts, requiring also that $n_j=1$.
\label{fig:phi1pihi2_ttbar}}
\end{center}
\end{figure}

\begin{figure}[!htbp]
\begin{center}
\includegraphics[height=0.4\textheight]{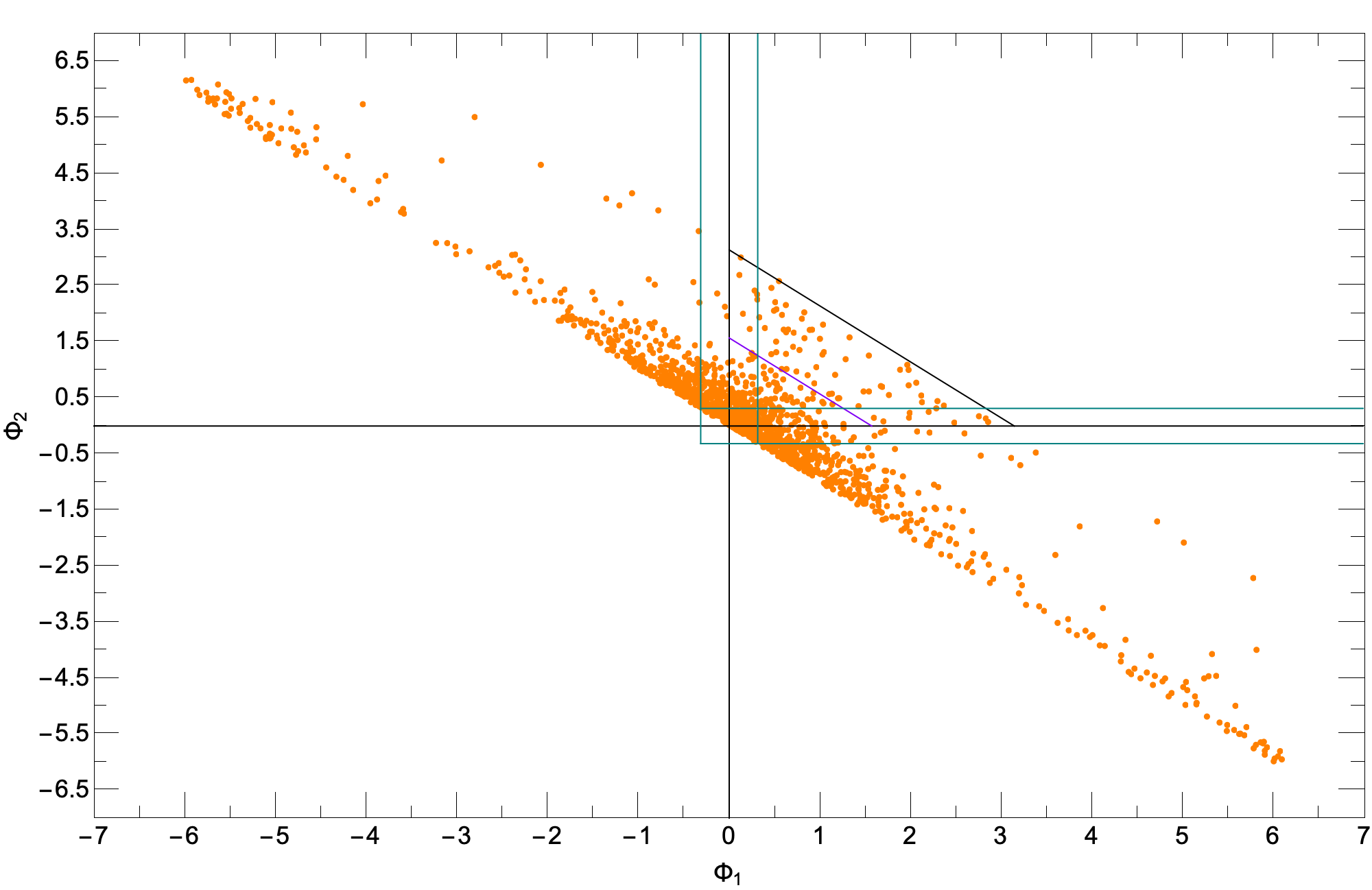}
\caption{Distribution in $\phi_1$ vs. $\phi_2$ plane for signal point 
BM1 after $C1$ cuts, requiring also that $n_j=1$.
\label{fig:phi1pihi2_BM1}}
\end{center}
\end{figure}

To characterize the $Z(\to \tau \bar{\tau})+j$ background, we show in
Fig.~\ref{fig:phi1pihi2_tautauj} a scatter plot of these events in the
$\phi_1\equiv \phi_{max}-\phi_{\hspace{1mm} \eslt}$ vs. $\phi_2\equiv
\phi_{\hspace{1mm} \eslt}-\phi_{min}$ plane. If the collinear
approximation for tau decays holds, we would expect that the $\tau\tau
j$ background selectively populates the top right quadrant with
$\phi_1>0$ and $\phi_2>0$ with $\phi_1+\phi_2 =\phi_{max}-\phi_{min} <
\pi$, and significantly smaller than $\pi$ when the tau pair emerges
with a small opening angle in the transverse plane. We see from the
figure that there is a small, but significant, spill-over into the
region where $\phi_1$ or $\phi_2$ assumes small negative values; {\it
  i.e.} where $\vec{\eslt}$ lies just outside the cone formed by
$\vec{\ell_1}$ and $\vec{\ell_2}$. This spill-over arises from
asymmetric decays of the $Z$ where one of the taus (the one emitted
backwards from the $Z$ direction) is relatively less relativistic so
that the collinear approximation works poorly, or because hadronic
energy mismeasurements skew the direction of $\vec{\eslt}$.
Indeed we see from Fig.~\ref{fig:phi1pihi2_tautauj} that the $\tau\tau
j$ background mostly populates the triangle in the top-right corner of
the $\phi_1$ vs. $\phi_2$ plane, and $\phi_1+\phi_2< f\pi$ where the
fraction $0<f<1$, with a spill-over into the strips where one of
$\phi_{1,2}$ is slightly negative.  For signal events and for the other
backgrounds, $\phi_{\hspace{1mm} \eslt}$ will be uncorrelated with $\phi_{min}$ and
$\phi_{max}$, and so their scatter plots will extend to the other
quadrants. This is illustrated for the $t\bar{t}$ background in
Fig.~\ref{fig:phi1pihi2_ttbar} and for signal point BM1 in
Fig.~\ref{fig:phi1pihi2_BM1}. In these cases, we indeed see a wide
spread in $\phi_1$ and $\phi_2$ between $\pm 2\pi$.

To efficiently veto the $\tau\bar{\tau}j$ background, we have examined nine cases
of angular cuts. To optimize the effect of the boost on the opening
angle of the two taus, we examine three ranges of $\phi_1+\phi_2$: 
\bi
\item {\it a1}: $\phi_1,\ \phi_2 >0$, 
\item {\it b1}: $\phi_1,\ \phi_2 >0$ with $\phi_1+\phi_2<\pi /2$, and 
\item {\it c1}: $\phi_1,\ \phi_2 >0$ with $\phi_1+\phi_2<2\pi /3$.  
\ei
Next, to optimize the width of the ``strip''
where the $\eslt$ vector is allowed to stray outside the cone formed by
the leptons, we also tried,
\bi
\item  {\it a2}, {\it b2} and {\it c2} where instead
$\phi_1,\ \phi_2> -\pi/10$, and 
\item {\it a3}, {\it b3} and {\it c3} with $\phi_1,\ \phi_2> -\pi/20$. 
\ei
The set which gives optimized
$S/\sqrt{BG(\tau\bar{\tau} j)}$ for LHC14 with 3000 fb$^{-1}$ was found to be set {\it
  b1}: 
\be {\rm veto\ the\ triangle}\ \ \ \phi_1,\ \phi_2> 0\ \ {\rm
  with}\ \ \phi_1+\phi_2<\pi /2, \label{eq:phi}
\ee 
along with an
additional veto of the $|\phi_1|$ and $|\phi_2|$ strips along the
positive $\phi_1$ and $\phi_2$ axes to further reduce background from
the spill-over of $\ \vec{\eslt}$ outside of the cone defined by the taus
that we already discussed:
%
%
\be {\rm
  strip\ cuts:\ veto} |\phi_{1,2}|<\pi /10 . \label{eq:strip}  
\ee 
We list signal and background rates after {\bf C1} cuts together with
the angle cuts  (\ref{eq:phi}) and (\ref{eq:strip}) in row 5 of
Table~\ref{tab:xsec}. In this case, we find that $\tau\bar{\tau}j$ background
is reduced from cut set {\bf C1} by a factor $\sim 43$ (compared to a factor
$\sim 4$ for the $m_{\tau\tau}^2< 0$ cut) whilst signal efficiency for 
the point BM1 is almost 60\% (compared to $\sim 75$\% for the $m_{\tau\tau}^2< 0$
cut).\footnote{The handful of events at values of $\phi_{1}$ or $\phi_2$
  close to $2\pi$ in Fig.~\ref{fig:phi1pihi2_tautauj} occurs for the
  same reason as events along the strips about $|\phi_{1,2}|\sim 0$;
  {\it e.g.} one lepton and $\eslt$ directions may be close to zero in
  azimuth, with the azimuthal angle of the other lepton being just under
  $2\pi$. These would be eliminated by amending the veto region in the
  strip cuts in Eq.~(\ref{eq:strip}) to be smaller than $\pi/10$ mod
  $2\pi$. This modification would further reduce the $\tau\bar{\tau}j$
  background listed in the row labeled {\bf C1} + angle by about a
  factor 2. We have not included this reduction in this analysis, but it is included in an updated report Ref. \cite{Baer:2022qrw}.}  We
also see that signal efficiency for the other two benchmark points is
nearly the same for the angular and for the $m_{\tau\tau}^2< 0$ cuts.
We regard the angular cuts as a significantly improved method for
reducing $\tau\bar{\tau}j$ background relative to signal. We note that the
other SM backgrounds are not as efficiently reduced by the angular cut
as by the $m_{\tau\tau}^2< 0$ cut, and it is with this in mind that we
turn to the examination of other distributions below.


\subsection{Additional distributions to reduce $t\bar{t}$, $WWj$ and
  other backgrounds} \label{subsec:c2cuts}

\begin{figure}[!htbp]
\begin{center}
\includegraphics[height=0.4\textheight]{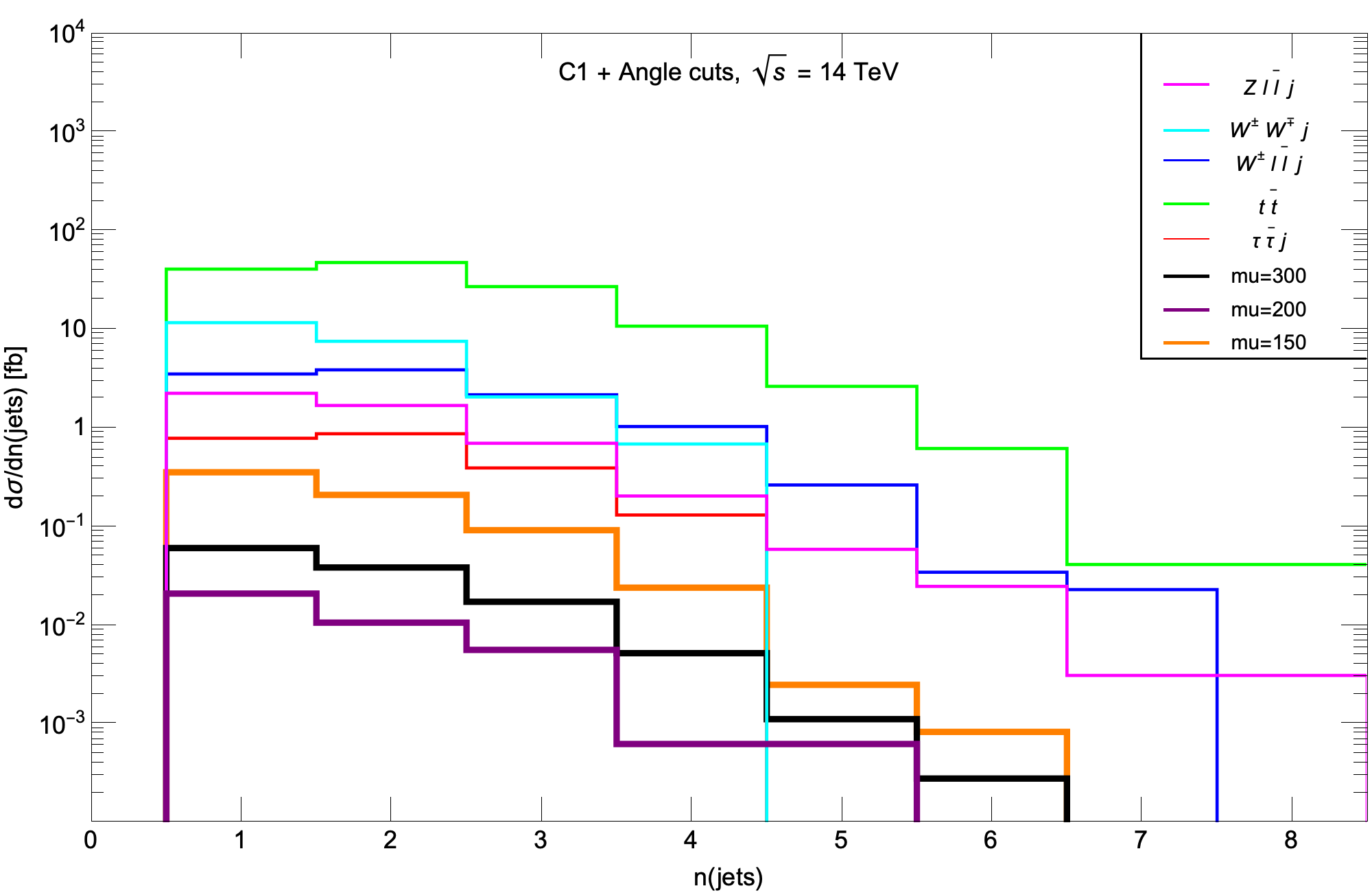}
\caption{Distribution in $n(jet)$ for three SUSY BM models with 
$\mu =150,\ 200$ and 300 GeV along with SM backgrounds after $C1$ and
  the angular cuts described in the text.
\label{fig:njets}}
\end{center}
\end{figure}

We have seen that after {\bf C1} cut set augmented by the angular cuts,
the main SM backgrounds arise from $t\bar{t}$ and $WWj$ production
followed by leptonic decays of the top and $W$-bosons.  Since $t\bar{t}$
production leads typically to events with two hard daughter $b$-quarks,
we begin with the examination of the jet multiplicity $n(jets)$ in
Fig. \ref{fig:njets}.  The signal distributions are shown as thick
orange, black and purple histograms for the benchmark cases, BM1, BM2 and BM3,
respectively, and they all feature steadily falling $n(jets)$
distribution since jets only arise from ISR.  In contrast, $n(jets)$
from $t\tbar$ production has a rather flat distribution out to
$n(jets)\sim 3$ with a steady drop-off thereafter.  The other EW
backgrounds also feature falling $n(jet)$ distributions.  Restricting
$n(jets)\sim 1-2$ should cut $t\tbar$ background substantially with
relatively small cost to signal.

\begin{figure}[!htbp]
\begin{center}
\includegraphics[height=0.4\textheight]{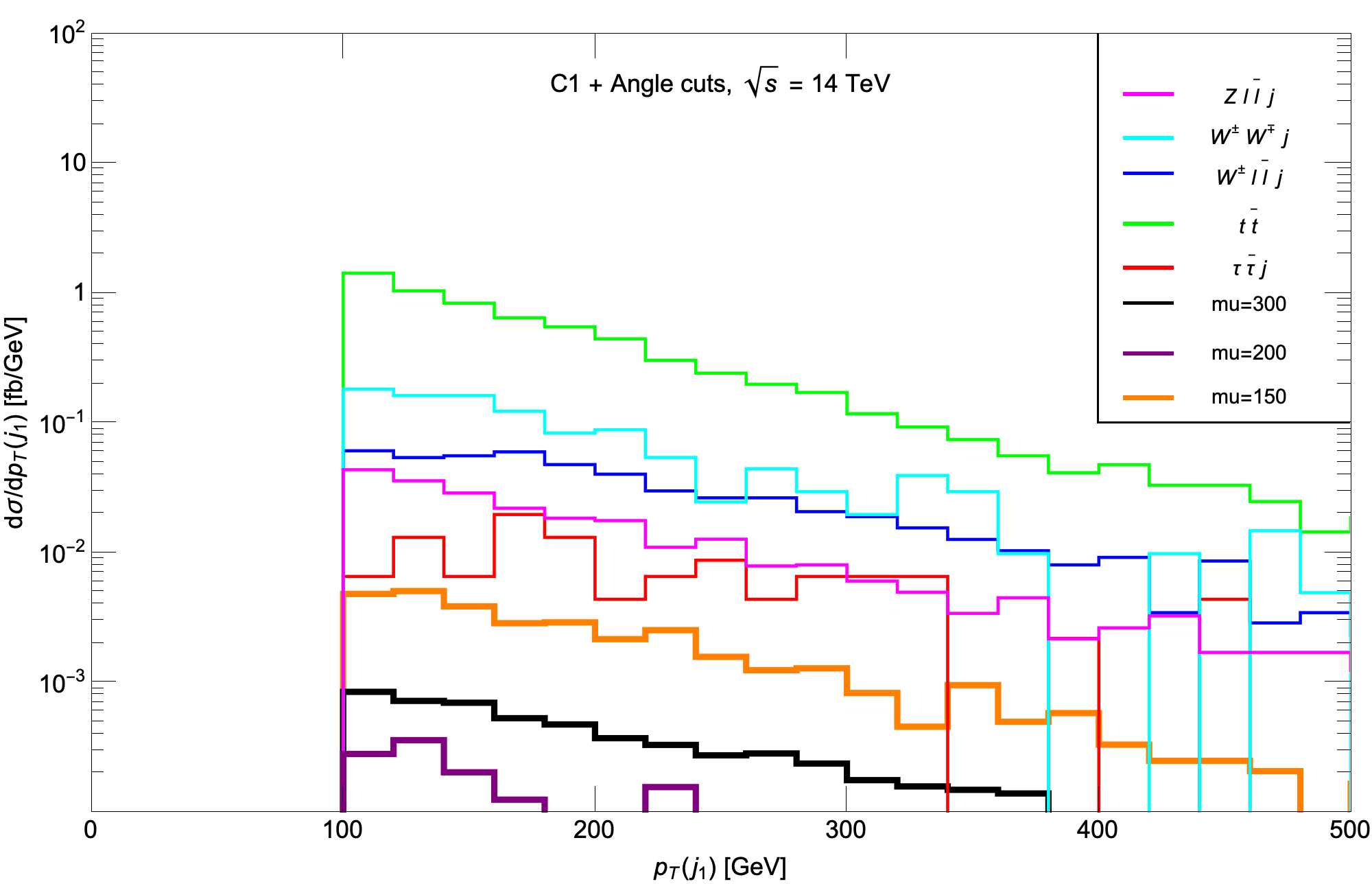}
\caption{Distribution of the hardest jet $p_T(j_1)$ for the 
three SUSY BM models with 
$\mu =150,\ 200$ and 300 GeV and for SM backgrounds after $C1$ and
angular cuts.
\label{fig:ptj1}}
\end{center}
\end{figure}

\begin{figure}[!htbp]
\begin{center}
\includegraphics[height=0.4\textheight]{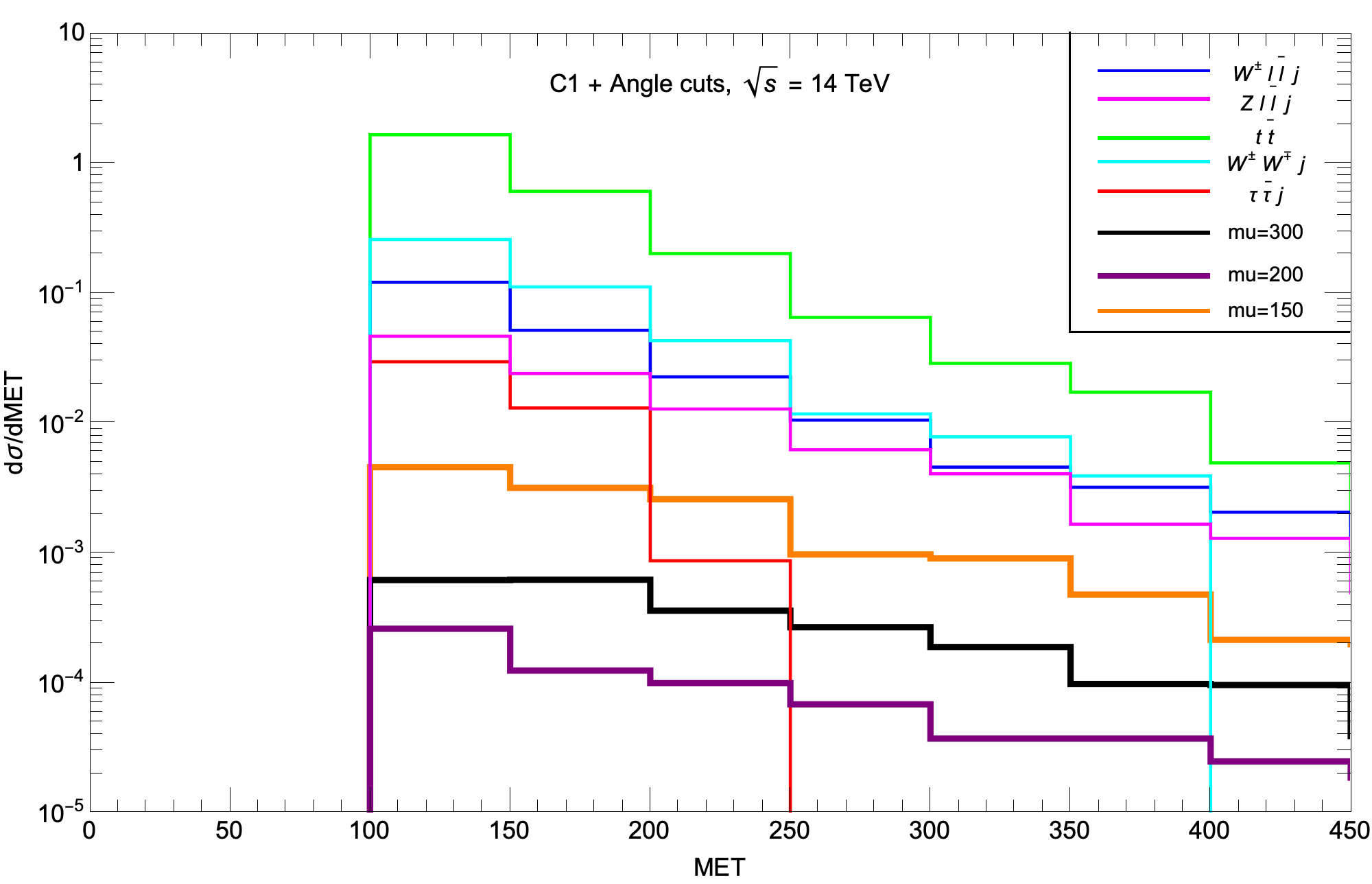}
\caption{Distribution of $\eslt$ for the three SUSY BM models with $\mu
  =150,\ 200$ and 300 GeV and for SM backgrounds after $C1$ and angular
  cuts.
\label{fig:eslt}}
\end{center}
\end{figure}

We continue our examination by showing in Fig. \ref{fig:ptj1} and
Fig.~\ref{fig:eslt} the distribution of the highest $p_T$ jet and of
$\eslt$, respectively, again after {\bf C1} and angular cuts.  We see
that both distributions are backed up against the cut and falling
steeply, for both the signal cases as well as for the backgrounds. While 
these distributions may be falling slightly faster for the top
background as compared to the signal, it is clear that requiring 
harder cuts on either $p_T(j_1)$ or $\eslt$ would greatly reduce the
already small signal.


\begin{figure}[!htbp]
\begin{center}
\includegraphics[height=0.4\textheight]{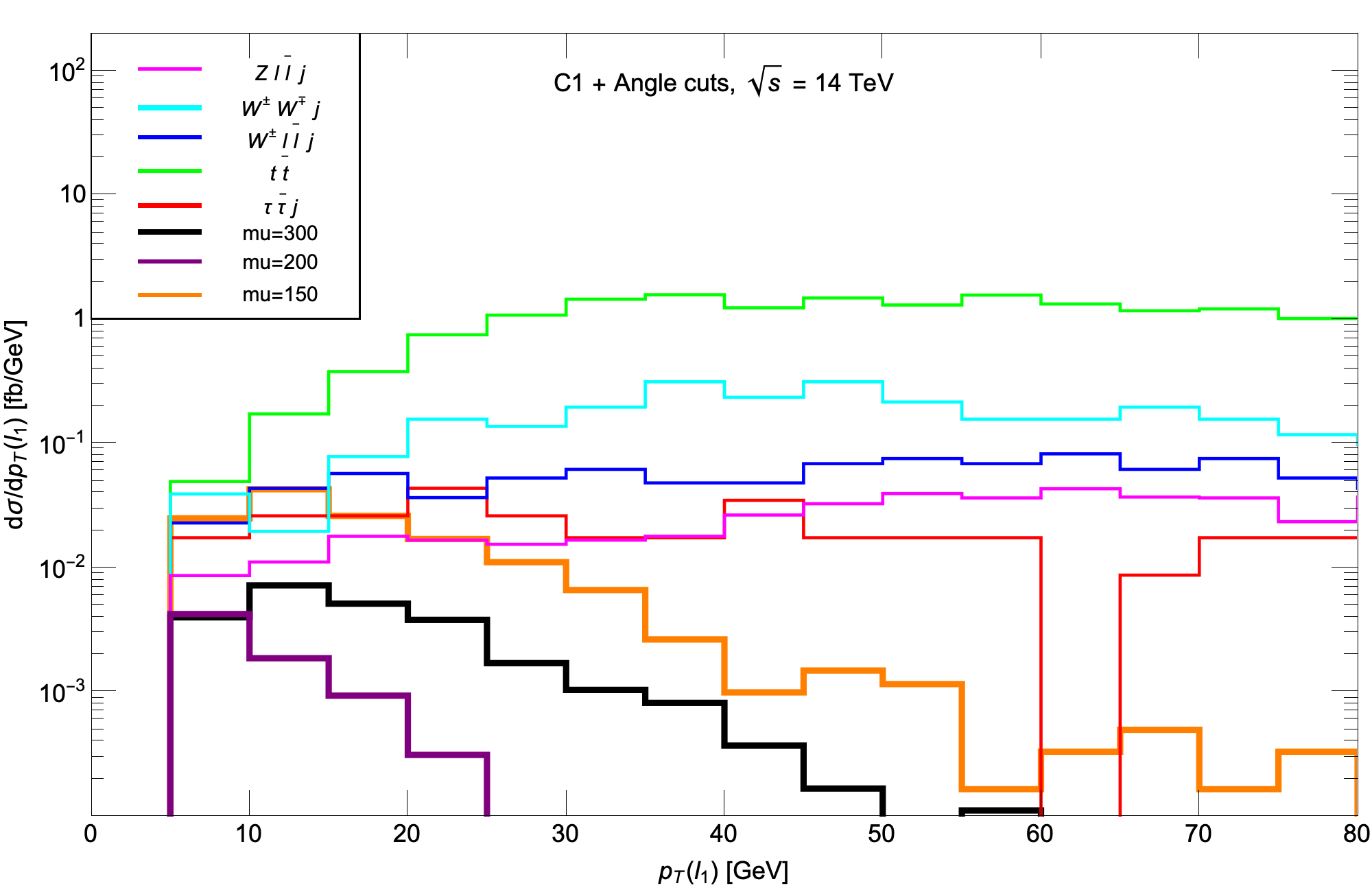}
\caption{Distribution of the transverse momentum of the hard lepton
  $p_T(\ell_1)$ for the three SUSY BM models with $\mu =150,\ 200$ and
  300 GeV and for SM backgrounds after $C1$ and the angular cuts.
\label{fig:ptl1}}
\end{center}
\end{figure}

Turning to the leptons in the events, we show in Fig. \ref{fig:ptl1} the
distributions in $p_T(\ell_1)$, the highest $p_T$ isolated
lepton. As expected, the signal distributions are very soft whereas the
corresponding distributions from $t\tbar$, $WWj$ (and even from the
residual $\tau\bar{\tau}j$ events) extend to far beyond where the signal
distributions have fallen to 10-20\% of their peak value.  In this case,
an upper bound on $p_T(\ell_1)\alt 25-40$ GeV might be warranted, at
least for SUSY signal cases where the neutralino mass gap is $\alt
20$~GeV.

\begin{figure}[!htbp]
\begin{center}
\includegraphics[height=0.4\textheight]{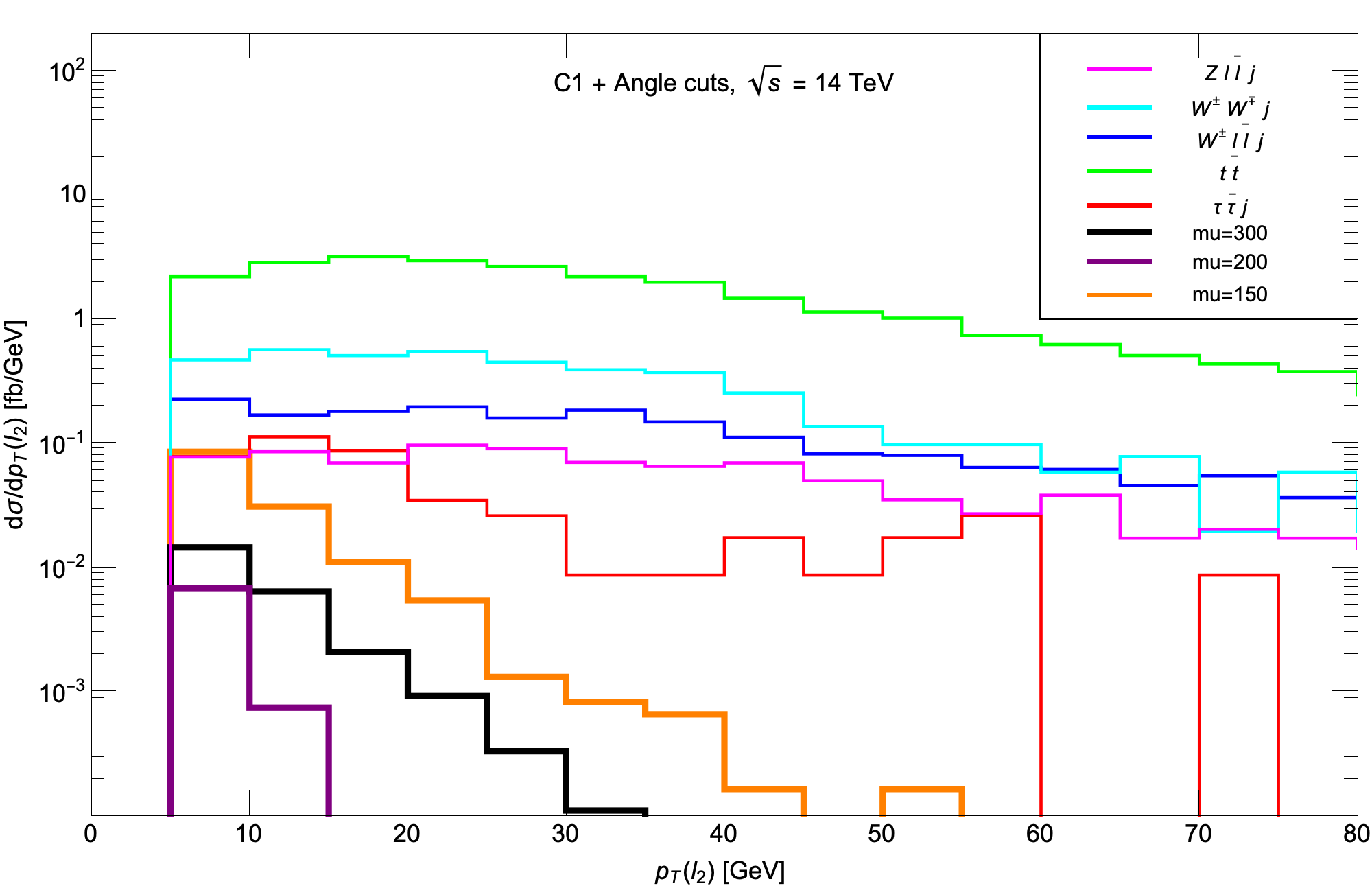}
\caption{Distribution of the softer lepton $p_T(\ell_2)$ for the three
  SUSY BM models with $\mu =150, 200$ and 300 GeV SUSY BM models and for SM
  backgrounds after $C1$ and angular cuts.
\label{fig:ptl2}}
\end{center}
\end{figure}

In Fig. \ref{fig:ptl2}, we show the resultant distributions in $p_T$ of the
lower $p_T$ isolated lepton. 
In this case, the three signal BM models have sharply falling distributions 
whilst many of the SM background distributions are rather flat 
out to high $p_T(\ell_2)$. Requiring $p_T(\ell_2):5-20$ GeV should save the 
bulk of signal events (at least as long as the neutralino 
mass gap is not very large) while rejecting the majority of the background.

\begin{figure}[!htbp]
\begin{center}
\includegraphics[height=0.4\textheight]{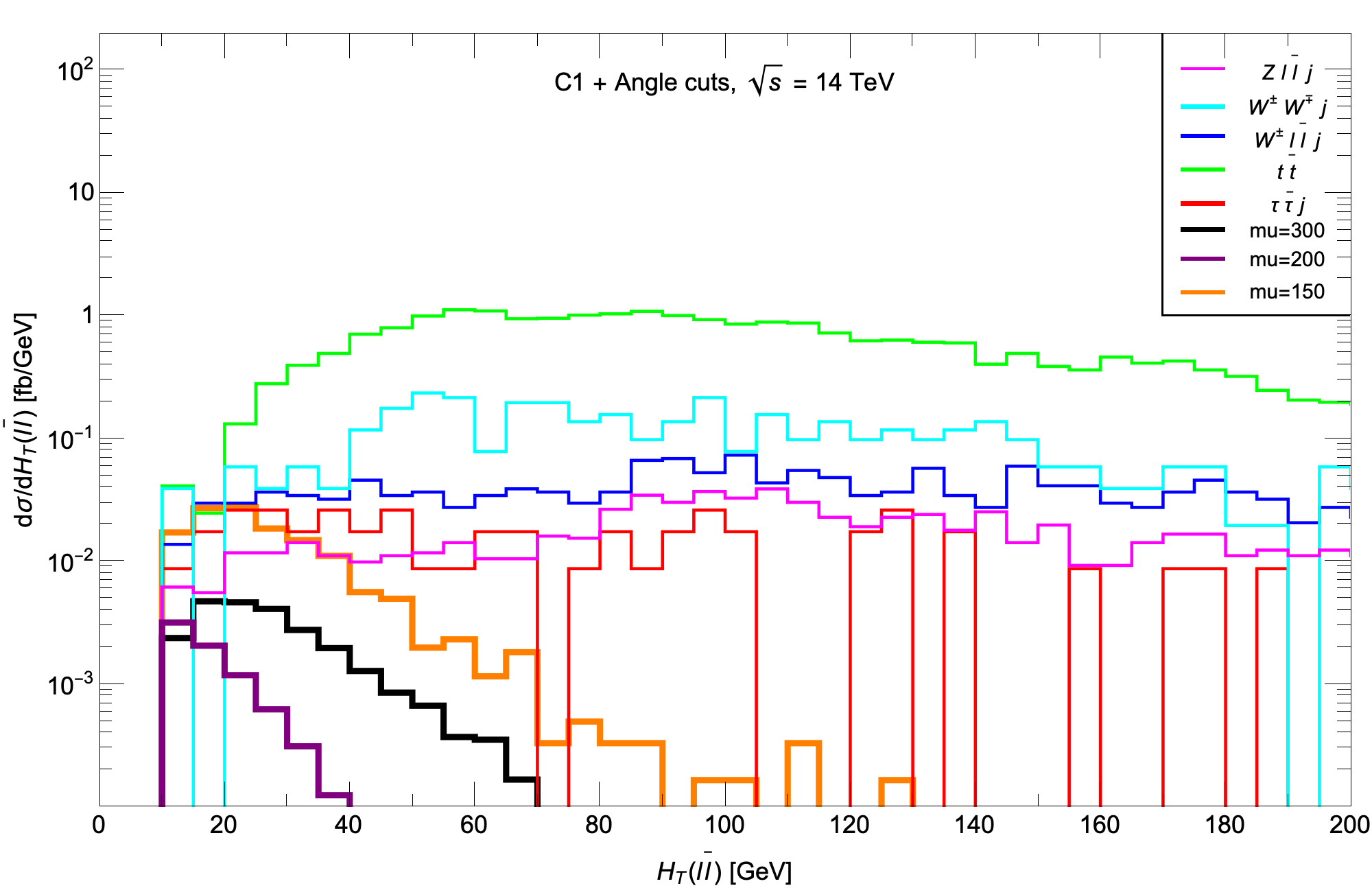}
\caption{Distribution in $H_T(\ell\bar{\ell})$ for the three SUSY BM
  models with $\mu =150$ GeV, 200~GeV and 300 GeV  and for SM
  backgrounds after $C1$ and angular cuts.
\label{fig:ht}}
\end{center}
\end{figure}

In Fig. \ref{fig:ht}, we plot the scalar sum of lepton $p_T$ values
$H_T(\ell\bar{\ell})\equiv |p_T(\ell_1)|+|p_T(\ell_2)|$.\footnote{The
  $H_T$ variable was originally introduced in Fig. 4 of
  Ref. \cite{Baer:1988kq} to help discriminate $t\tbar$ signal events
  from $W+jets$ background in the Tevatron top-quark searches.} Since
signal gives rise to soft OS/SF dileptons while most backgrounds have at
least one hard lepton, then we expect harder $H_T$ distributions from
background.  The figure illustrates that this is indeed the case, and
that a cut $H_T(\ell\bar{\ell})\alt 50-60$ GeV would enhance the signal
relative to the background. Of course, $|p_T(\ell_1)|, |p_T(\ell_2)|$ and
$H_T$ are strongly correlated, so that cutting on any two of these would
serve for our purpose.

\begin{figure}[!htbp]
\begin{center}
\includegraphics[height=0.4\textheight]{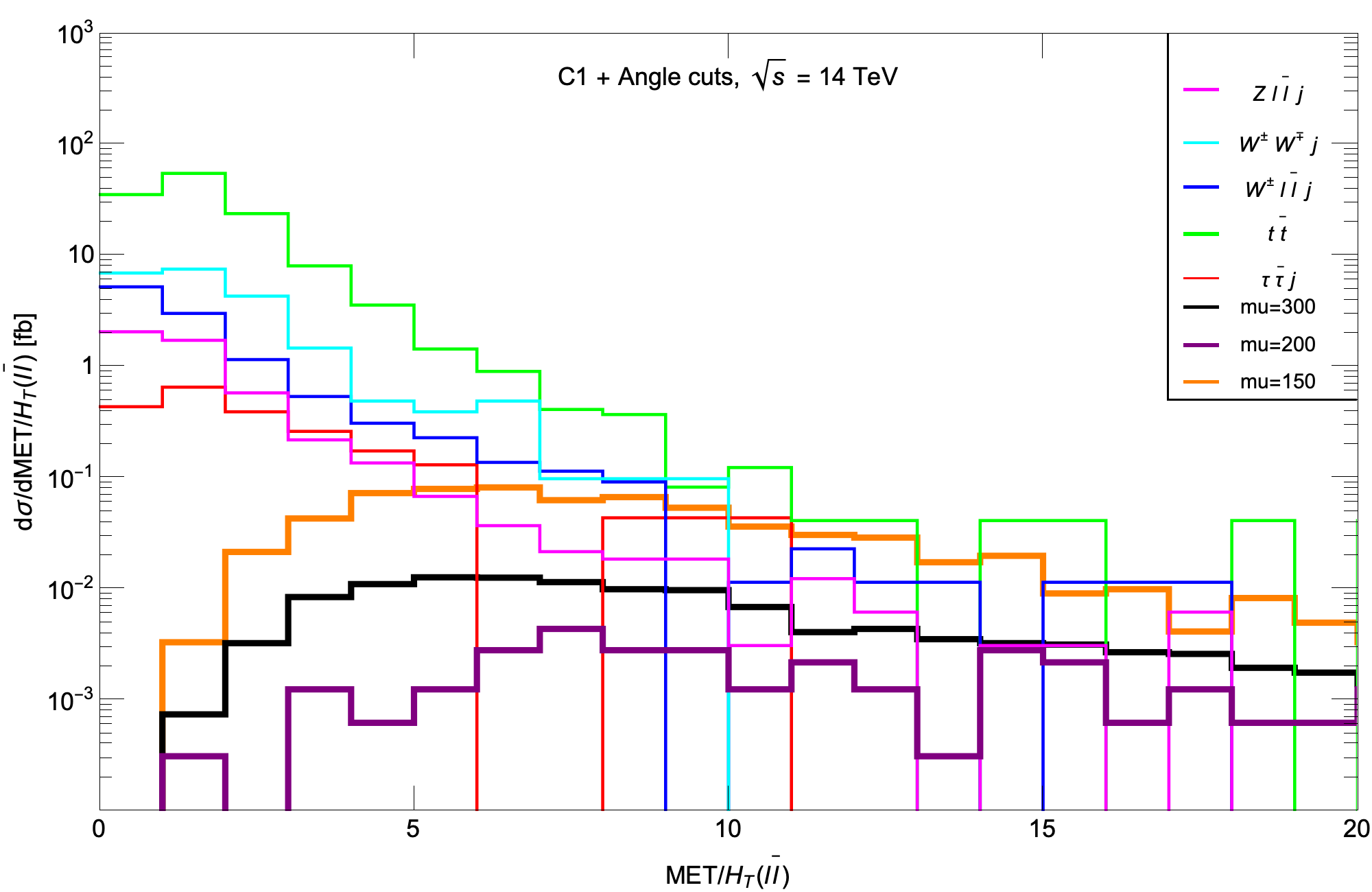}
\caption{Distribution of $\eslt /H_T(\ell)$ for three SUSY BM models with 
$\mu =150,\ 200$ and 300 GeV and for  SM backgrounds after $C1$ cuts and
  angular cuts.
\label{fig:etht}}
\end{center}
\end{figure}

The distribution in $\eslt /H_T(\ell\bar{\ell})$ was found by the ATLAS
collaboration to be an effective signal-to-background discriminator in
Ref. \cite{atlas1}. The signal is expected to exhibit a soft $H_T$
distribution compared to a hard $\eslt$ distribution from recoil of SUSY
particles against the ISR jet. Thus, signal is expected to exhibit a
hard $\eslt /H_T$ distribution compared to background. In
Fig. \ref{fig:etht}, we show the relevant SUSY BM distributions along
with SM backgrounds.  Indeed, almost all $t\tbar$ events -- and also
most other events -- lie with $\eslt /H_T\alt 4$ while signal events
peak around $\eslt /H_T\sim 5-10$.  We will, in addition, require $\eslt
/H_T >4$ for our next cut set {\bf C2}.

\subsection{$C2$ cuts: signal, BG and distributions}
\label{ssec:C2}

In light of the distributions just discussed, we next include the
following cut set {\bf C2} to enhance the higgsino signal over top,
$WWj$ and the other EW backgrounds: 
\bi
\item the cut set {\bf C1} together with the ${\rm angle\ cuts}$,
\item $n(jets) = 1$,
\item $p_T(\ell_2 ):5-15$ GeV,
\item $H_T(\ell\bar{\ell})<60$ GeV
\item $\eslt/H_T(\ell\bar{\ell} )>4$, and
\item $m(\ell\bar{\ell})<50$ GeV.
\ei 

The reader will have noticed that we have included an upper limit on the
invariant mass of the dilepton pair. This cut is motivated from the fact
that the invariant mass distributions of dileptons from $\tz_2\to
\tz_1\ell\bar{\ell}$ decay is kinematically bounded by
$m_{\tz_2}-m_{\tz_1}$, and further that leptons from the decays of {\em
  different} charginos/neutralinos also tend to have small energies (and
hence also small $m(\ell\bar{\ell})$) because the higgsino spectrum is
compressed. In contrast, leptons from decays of background tops and
$W$-bosons tend to be hard (see Fig.~\ref{fig:ptl1} and
Fig.~\ref{fig:ptl2}) and, because the lepton directions are
uncorrelated, the corresponding background dilepton mass distributions
are relatively flat out to very large values of $m(\ell\bar{\ell})$. Although
we do not show it, we have checked that the requirement $m(\ell\bar{\ell})<
50$~GeV, efficiently reduces much of the background while retaining most
of the higgsino signal as long as the higgsino spectrum is compressed.

We see from the penultimate row of Table~\ref{tab:xsec} that after {\bf
  C2} cuts, the leading $t\bar{t}$
background has dropped by a factor $\sim 130$, and the total SM
background has dropped to $\sim 1.1$\%, while the signal is
retained with an efficiency of 40-60\%.  At this point, the total
background is just below 2~fb.
Clearly, the signal cross
section is small, and the large integrated luminosities expected at the
HL-LHC will be necessary for the detection of the signal if the higgsino
mass is close to its naturalness bound of 300-350~GeV, or if the
higgsino spectrum is maximally compressed, consistent with naturalness.

\begin{figure}[!htbp]
\begin{center}
\includegraphics[height=0.4\textheight]{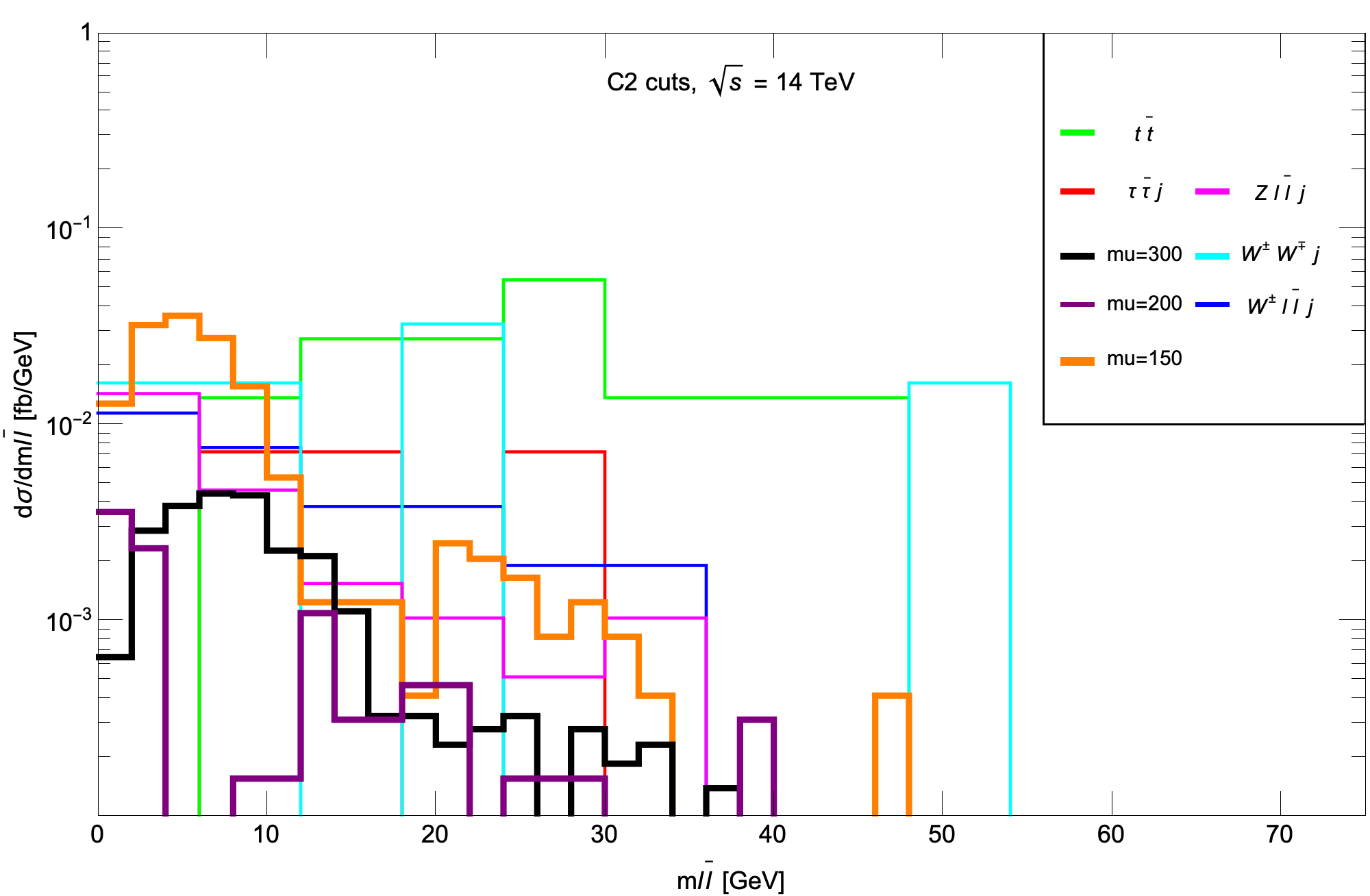}
\caption{Distribution in $m(\ell\bar{\ell})$ for the three SUSY BM models with 
$\mu =150,\ 200$ and 300 GeV, and for SM backgrounds after $C2$ cuts.
\label{fig:mll}}
\end{center}
\end{figure}

To characterize the signal events, and further improve the
discrimination of the signal {\it vis-a-vis} the background, we examine
other distributions after {\bf C2} cuts, starting with the dilepton
invariant mass distribution in Fig.~\ref{fig:mll}. We can gauge that the SM
background distribution, summed over the backgrounds, is essentially
flat. In contrast, the signal distributions show an accumulation 
of events below $m_{\tz_2}-m_{\tz_1}$ together with a long tail (with 
a much smaller number of events) where
the two leptons originate in {\em different} charginos/neutralinos. 

\begin{figure}[!htbp]
\begin{center}
\includegraphics[height=0.4\textheight]{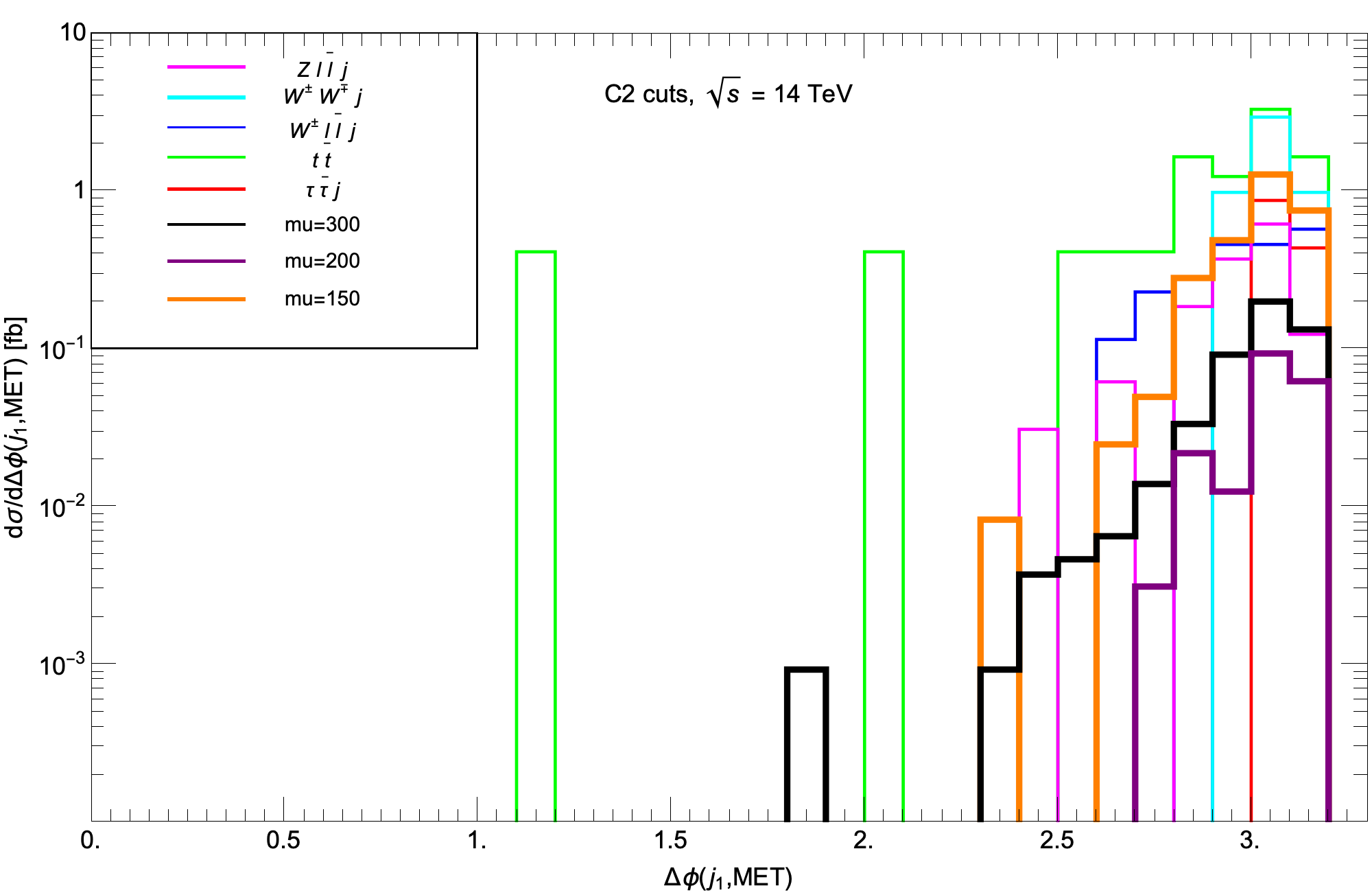}
\caption{Distribution in $\Delta\phi (jet,\eslt )$ for the three SUSY BM
  models with $\mu =150,\ 200$ and 300 GeV, and for SM backgrounds after
  $C2$ cuts.
\label{fig:dphi}}
\end{center}
\end{figure}

In Fig. \ref{fig:dphi}, we show the distribution in transverse
opening angle $\Delta\phi (j_1,\vec{\eslt})$. For the signal, where
the SUSY particles recoil strongly against the ISR jet, we expect
nearly back-to-back $\vec{p}_T(jet)$ and $\vec{\eslt}$ vectors.
This correlation is  expected to be somewhat weaker
from the $W\ell\bar{\ell}j$ 
and especially $t\bar{t}$  
backgrounds because these intrinsically contain additional activity
from decay products that do not form jets or identified leptons.  
Indeed, requiring 
$\Delta\phi (\vec{p}_T(j_1),\vec{\eslt})\agt 2$ appears to
give only a slight improvement in the signal-to-background ratio.
%

\begin{figure}[!htbp]
\begin{center}
\includegraphics[height=0.4\textheight]{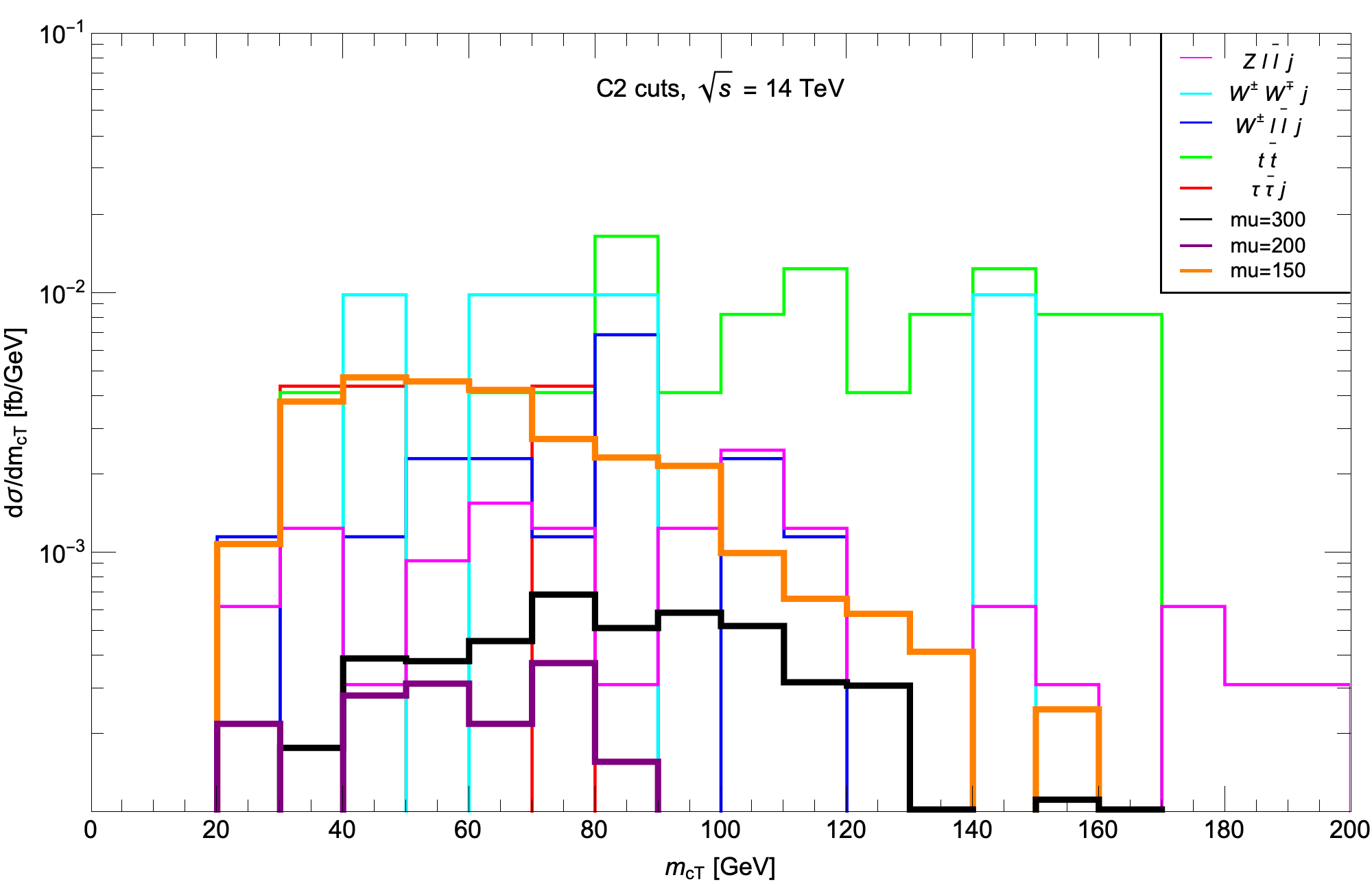}
\caption{Distribution in $m_{cT}(\ell^+\ell^-,\eslt )$ for the three SUSY BM
  models with $\mu =150,\ 200$ and 300 GeV, and for SM backgrounds
  after $C2$ cuts.
\label{fig:mct}}
\end{center}
\end{figure}

In Fig. \ref{fig:mct}, we plot the dilepton-plus-$\eslt$ cluster transverse
mass $m_{cT}(\ell\bar{\ell},\eslt)$. From the frame, we see the signal
distributions all have broad peaks around 20-100 GeV while several of
the backgrounds that contain harder leptons extend to well past 100 GeV.
Thus, a candidate analysis cut might include $m_{cT}\alt 100$ GeV.

\begin{figure}[!htbp]
\begin{center}
\includegraphics[height=0.4\textheight]{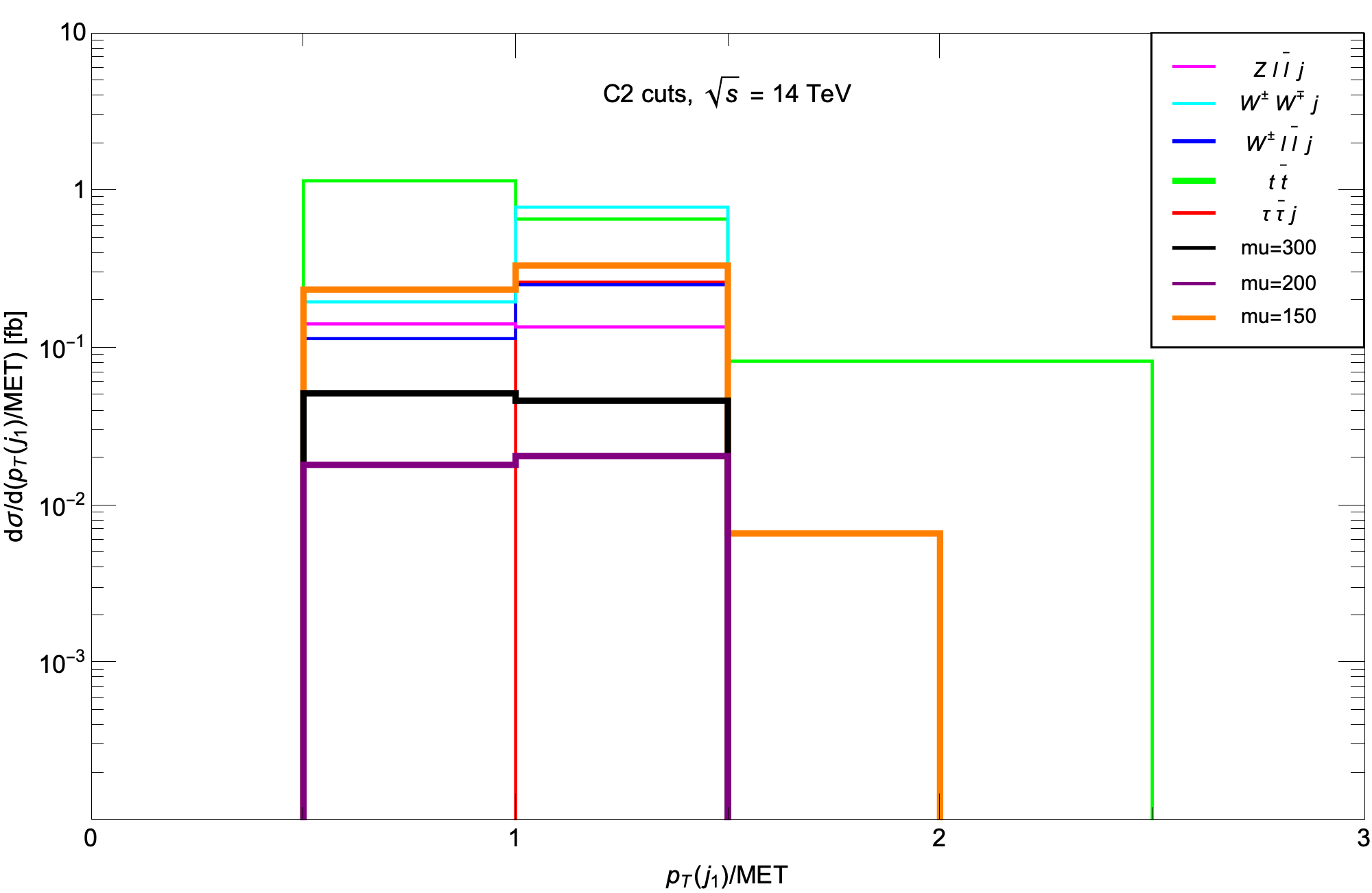}
\caption{Distribution in $E_T(jet)/ \eslt $ for three SUSY BM models with 
$\mu =150,\ 200$ and 300 GeV along with SM backgrounds after $C2$ cuts.
\label{fig:jetptbymet}}
\end{center}
\end{figure}

In Fig. \ref{fig:jetptbymet}, we plot the distribution in 
$p_T(j_1)/\eslt$. For the signal, we expect $\vec{\eslt}$ to mainly recoil
against the hard ISR jet so that signal would peak around $\sim 1$ 
since the dileptons are soft. In contrast, some of the backgrounds
will include harder high-$p_T$ objects so this ratio is expected to be 
less correlated. While both signal and BGs peak around $p_T(j_1)/\eslt\sim 1$, 
we note that several BG distributions extend out to
$p_T(j_1)/\eslt\sim 3$. Thus, we could require $p_T(j_1)/\eslt\alt 1.5$.

\begin{figure}[!htbp]
\begin{center}
\includegraphics[height=0.4\textheight]{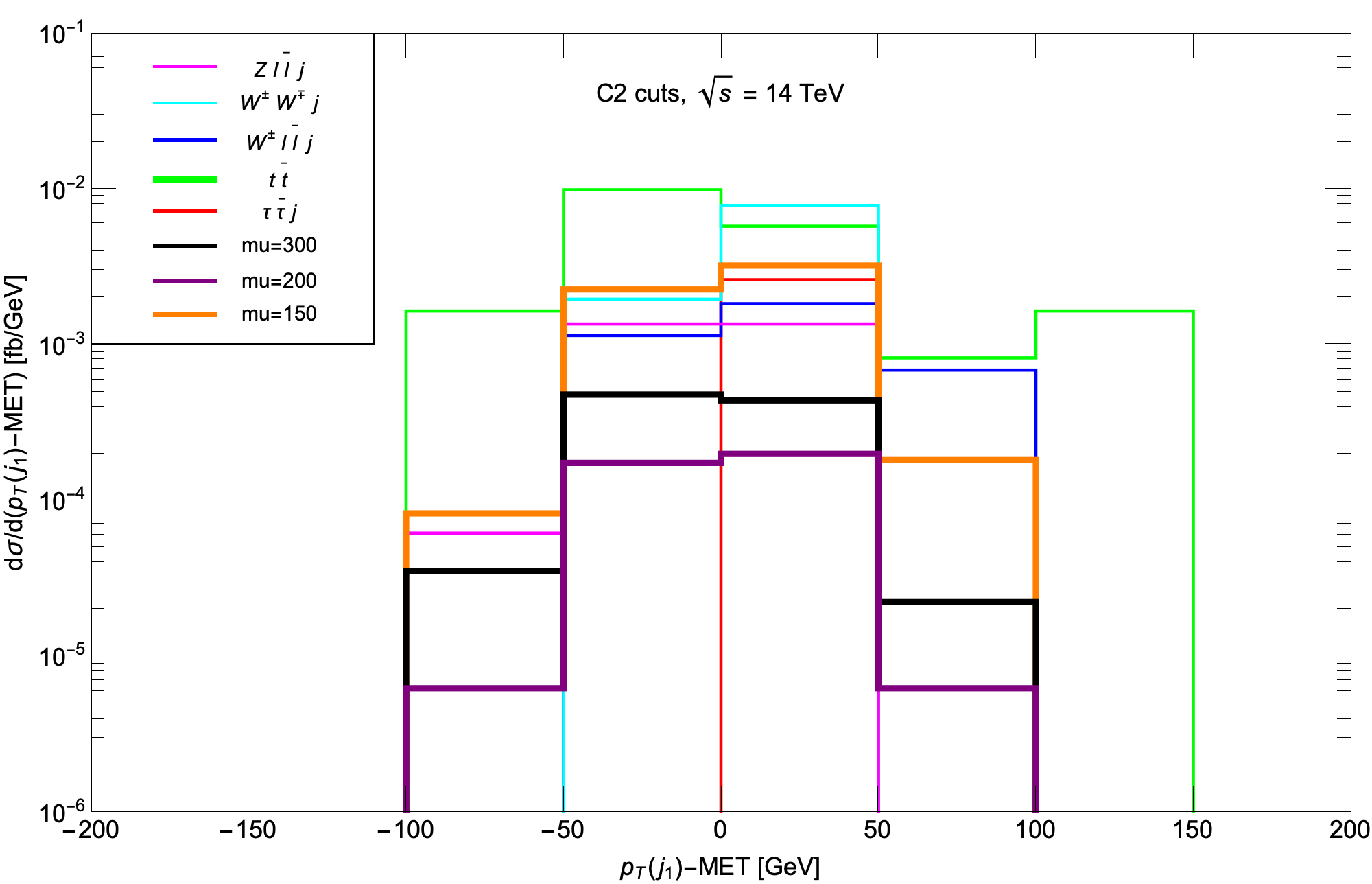}
\caption{Distribution in $E_T(jet)- \eslt$ for the three SUSY BM models with 
$\mu =150,\ 200$ and 300 GeV, and for SM backgrounds after $C2$ cuts.
\label{fig:ptjetminusmet}}
\end{center}
\end{figure}

A related distribution is to plot $p_T(j_1)-\eslt$, where again
signal values of $p_T(j_1)$ and $\eslt$ are expected to be nearly 
equal and opposite and so should peak around $\sim 0$. The backgrounds
have a similar peak structure, but extend to higher values 
especially in the positive direction. Therefore, we might require
$|p_T(j_1)-\eslt |\alt 100$ GeV. We note though that the considerations
in Figs.~\ref{fig:dphi}, \ref{fig:jetptbymet} and \ref{fig:ptjetminusmet}
have the same underlying physics, and hence the corresponding cuts 
are certainly correlated.

\begin{figure}[!htbp]
\begin{center}
\includegraphics[height=0.4\textheight]{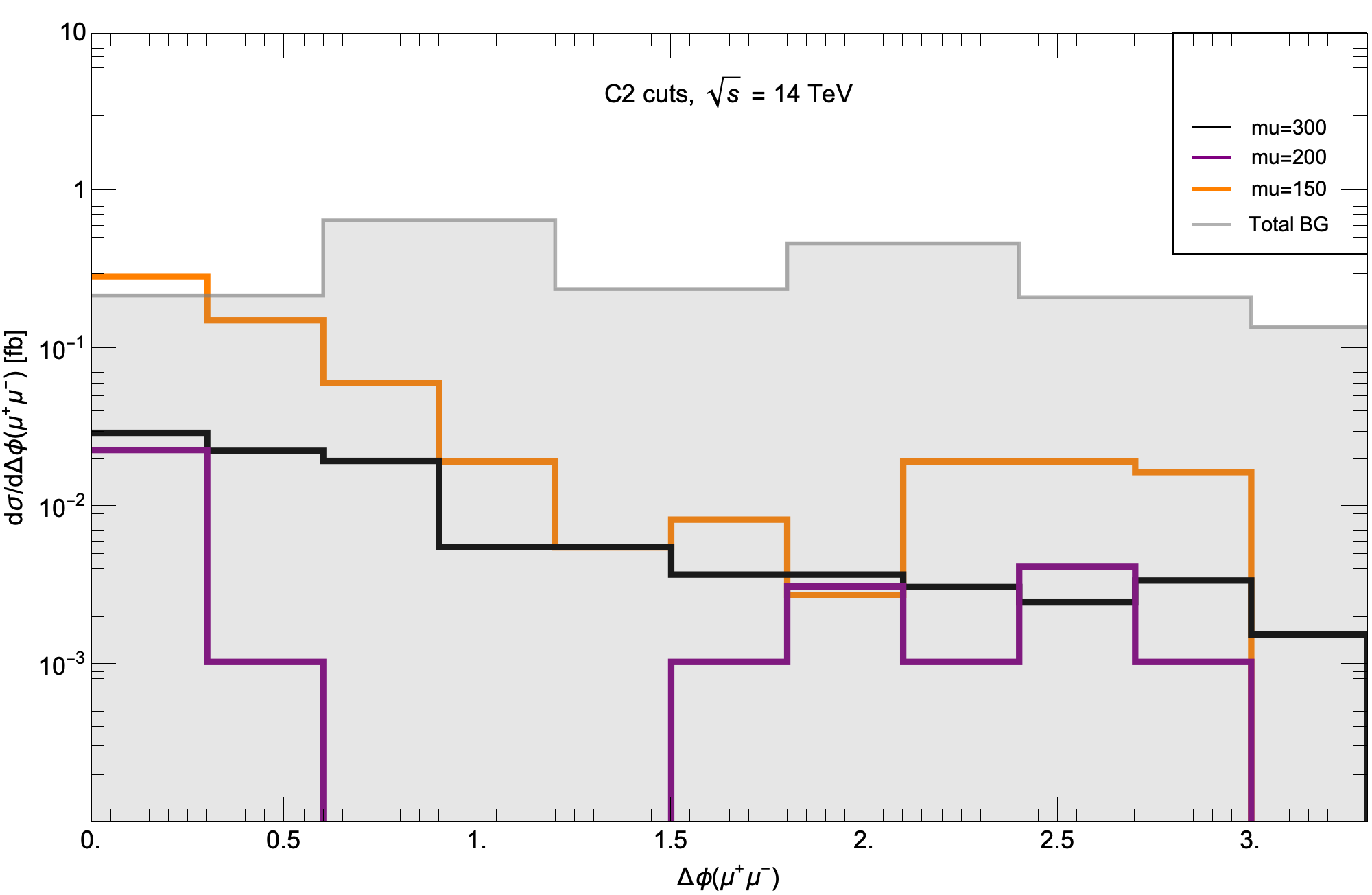}
\caption{Distribution in $\Delta\phi(\mu\bar{\mu} )$ for three SUSY BM models with 
$\mu =150,\ 200$ and 300 GeV along with SM backgrounds after $C3$ cuts.
\label{fig:delrmumu}}
\end{center}
\end{figure}

In Fig. \ref{fig:delrmumu}, we show the distribution in dimuon 
transverse opening angle $\Delta\phi (\mu\bar{\mu})$. 
In the signal case, we expect a significant recoil of $\tz_2$ from the ISR jet 
so that the muon pair originating from the $\tz_2\to \tz_1\mu\bar{\mu}$ decay
should be tightly collimated with small opening angle \cite{c_han}. For
the background processes, or for that matter from higgsino pair
production processes, where the leptons originate from different
particles or higher energy release decays, we do not expect the dilepton
pair to be so collimated, and indeed the total background is (within
fluctuations in our simulation) consistent
with being roughly flat in $\Delta\phi(\mu\bar{\mu})$.  Indeed, from the
figure we see that $\Delta\phi (\mu\bar{\mu})\sim 0-1$ for signal
processes while the SM BG processes tend to have opening angles less
well collimated and extending well past $\Delta\phi\sim 1.5$. Although
we have focussed on dimuons here, exactly the same consideration would
also apply to $e^+e^- +j+\eslt$ events, as long as the direction of the
electrons can be reliably measured.

In light of the above distributions, we next include the following
cut set {\bf C3} that includes:
\bi
\item all {\bf C2} cuts, 
\item $\Delta\phi (j_1,\eslt )>2.0$
\item $m_{cT}(\ell\bar{\ell} ,\eslt)<100$ GeV
\item $p_T(j_1)/\eslt <1.5$
\item $|p_T(j_1)-\eslt |<100$ GeV
\ei 
                     
\begin{figure}[!htbp]
\begin{center}
\includegraphics[height=0.3\textheight]{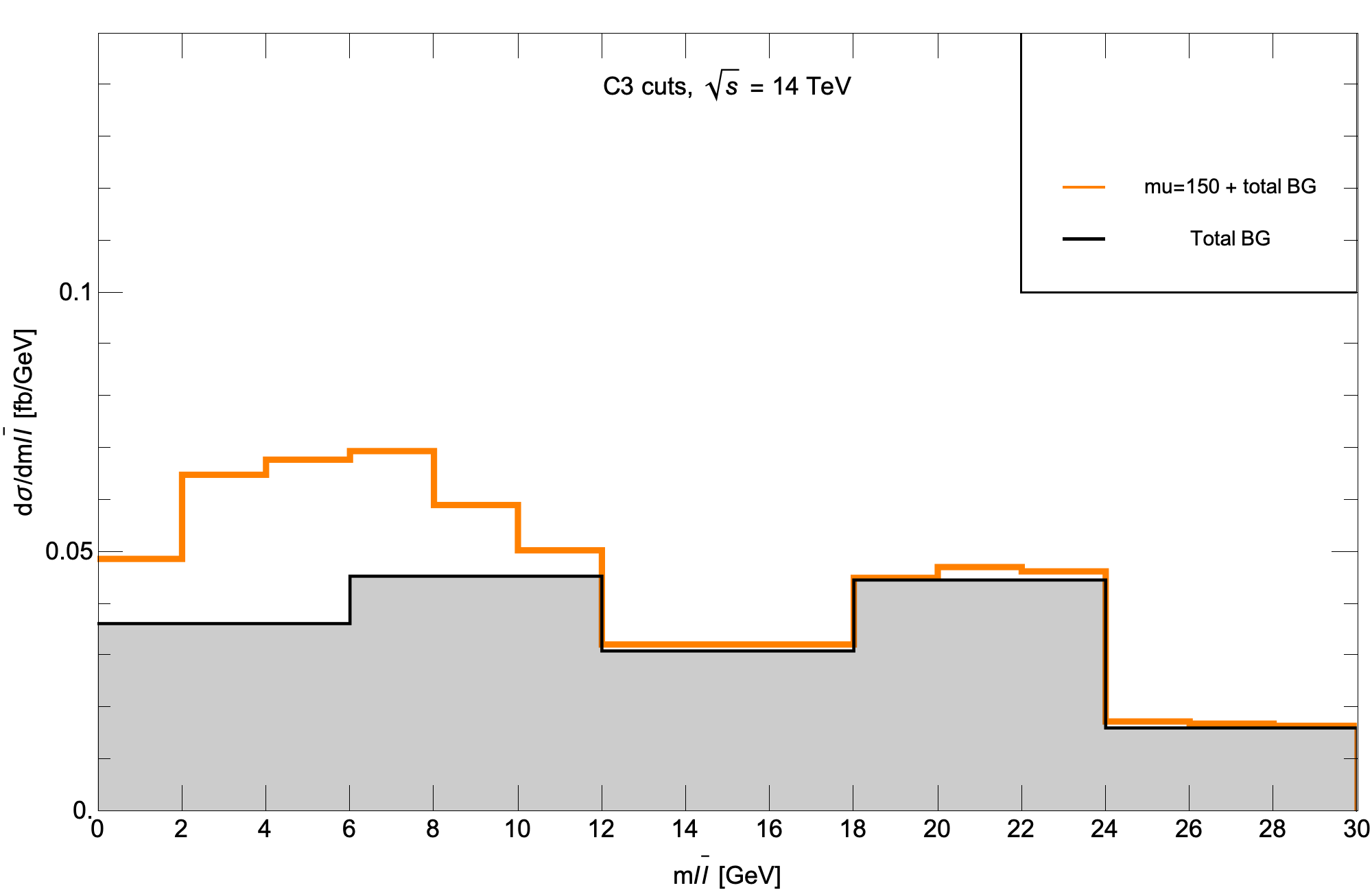}\\
\includegraphics[height=0.3\textheight]{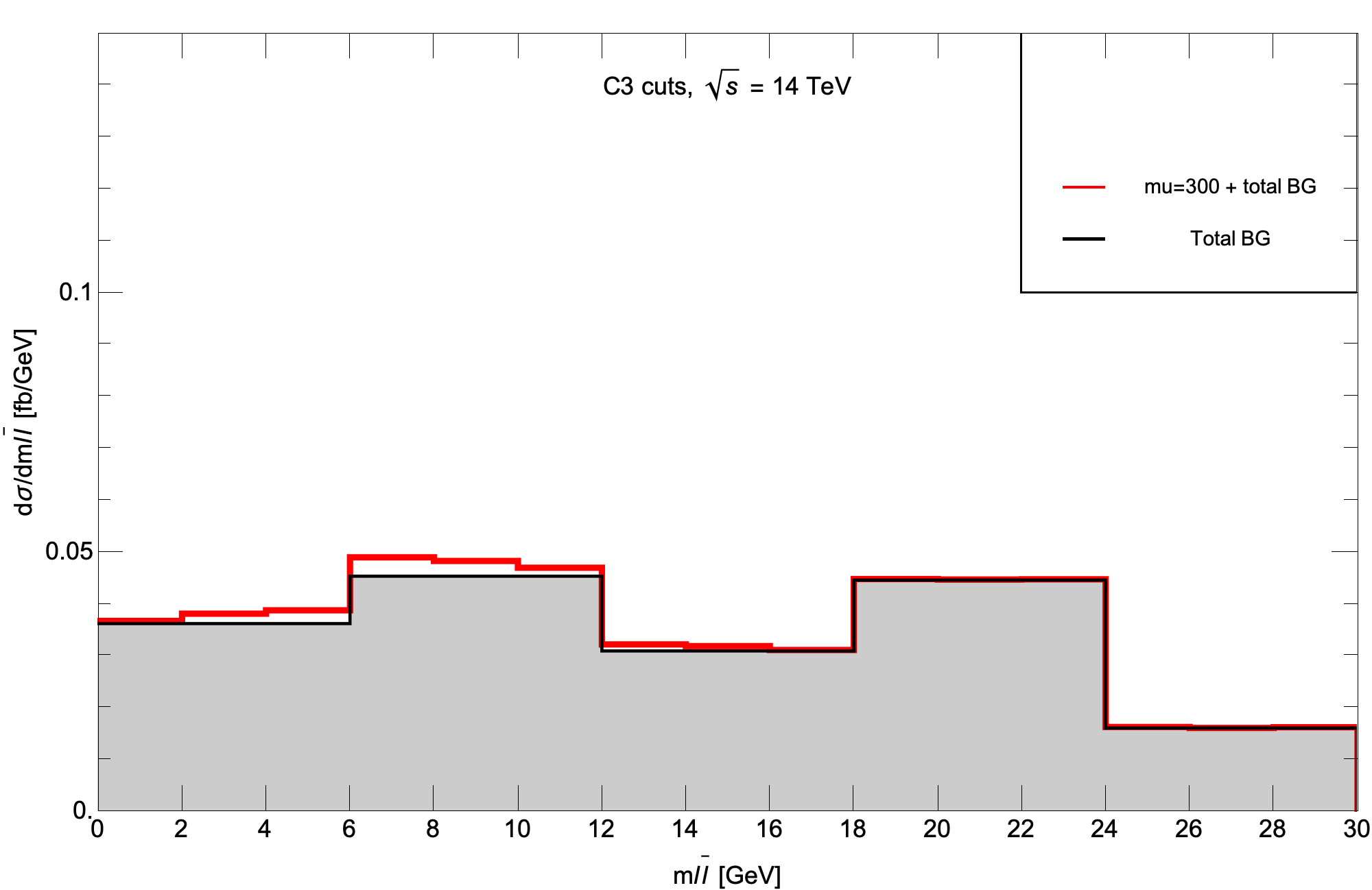}\\
\includegraphics[height=0.3\textheight]{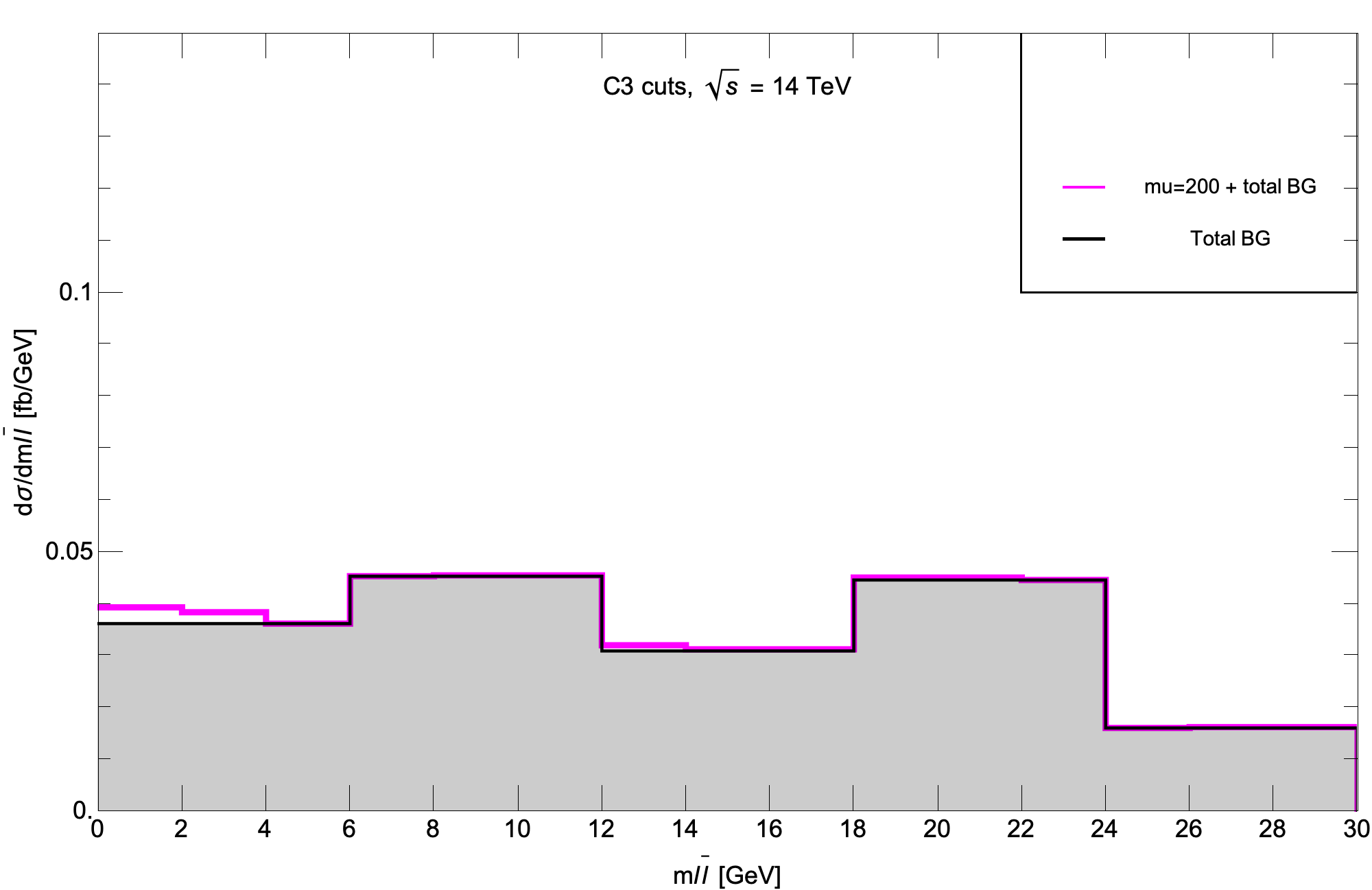}
\caption{Distribution of $m(\ell^+\ell^-)$ for the three SUSY BM models with 
$\mu =150,\ 300$ and 200 GeV, and for the  SM backgrounds after $C3$ cuts.
\label{fig:mllC3}}
\end{center}
\end{figure}

The OS/SF dilepton invariant mass after these {\bf C3} cuts is shown in
Fig. \ref{fig:mllC3}, this time on a linear scale.  The total background
is shown in gray, whilst signal-plus-background is the colored
histogram, and correspond to {\it a}) BM1 with $\Delta m=12$ GeV, {\it
  b}) BM2 with $\Delta m=16$ GeV and {\it c}) BM3 with $\Delta m=4.3$
GeV.  The idea here is to look for systematic deviations from SM
background predictions in the lowest $m(\ell\bar{\ell})$ bins.  Those
bins with a notable excess could determine the kinematic limit
$m(\ell\bar{\ell})<m_{\tz_2}-m_{\tz_1}$. By taking only the bins with a
notable excess, {\it i.e.}  $m(\ell\bar{\ell})<m_{\tz_2}-m_{\tz_1}$,
then it is possible to compute the cut-and-count excess above expected
background to determine a $5\sigma$ or a 95\% CL limit. The shape of the
distribution of the excess below the $\tz_2 \to \tz_1\ell\bar{\ell}$ end
point depends on the {\em relative sign} of the lighter neutralino
eigenvalues (these have opposite signs for higginos) and so could serve
to check the consistency of higgsinos as the origin of the
signal\cite{shape}. Of the three cases shown, this would be possible at
the HL-LHC only for the point BM1, since the tiny signal to background ratio
precludes the possibility of determining the signal shape in the other
two cases.

\section{LHC reach for higgsinos with 300-3000~fb$^{\bf {-1}}$}
\label{sec:reach}

In light of the above distributions, we next include the following
cut set {\bf C4}:
\bi
\item apply all {\bf C3} cuts,
\item then, require $m(\ell\bar{\ell})<m_{\tz_2}-m_{\tz_1}$.
\ei 

The reader could legitimately ask how we could implement this since we
do not {\em a priori} know the neutralino mass gap. The location of the
mass gap can be visually seen for BM1, but would be obscured by the
background for the other two cases. What we really mean is to measure
the cross section with $m_{\ell\ell}< m_{\ell\ell}^{\rm cut}$, varying
the value of $m_{\ell\ell}^{\rm cut}$ and looking for a rise in the (low
mass) region where events from $\tz_2\to\tz_1\ell\bar{\ell}$ would be
expected to accumulate. In the following, we will assume that once we
have the data, the region where the higgsino signal is beginning to
accumulate will be self-evident.

\begin{figure}[!htbp]
\centering
\begin{subfigure}[t]{0.44\textwidth}
  \centering
  \includegraphics[width=1.1\linewidth]{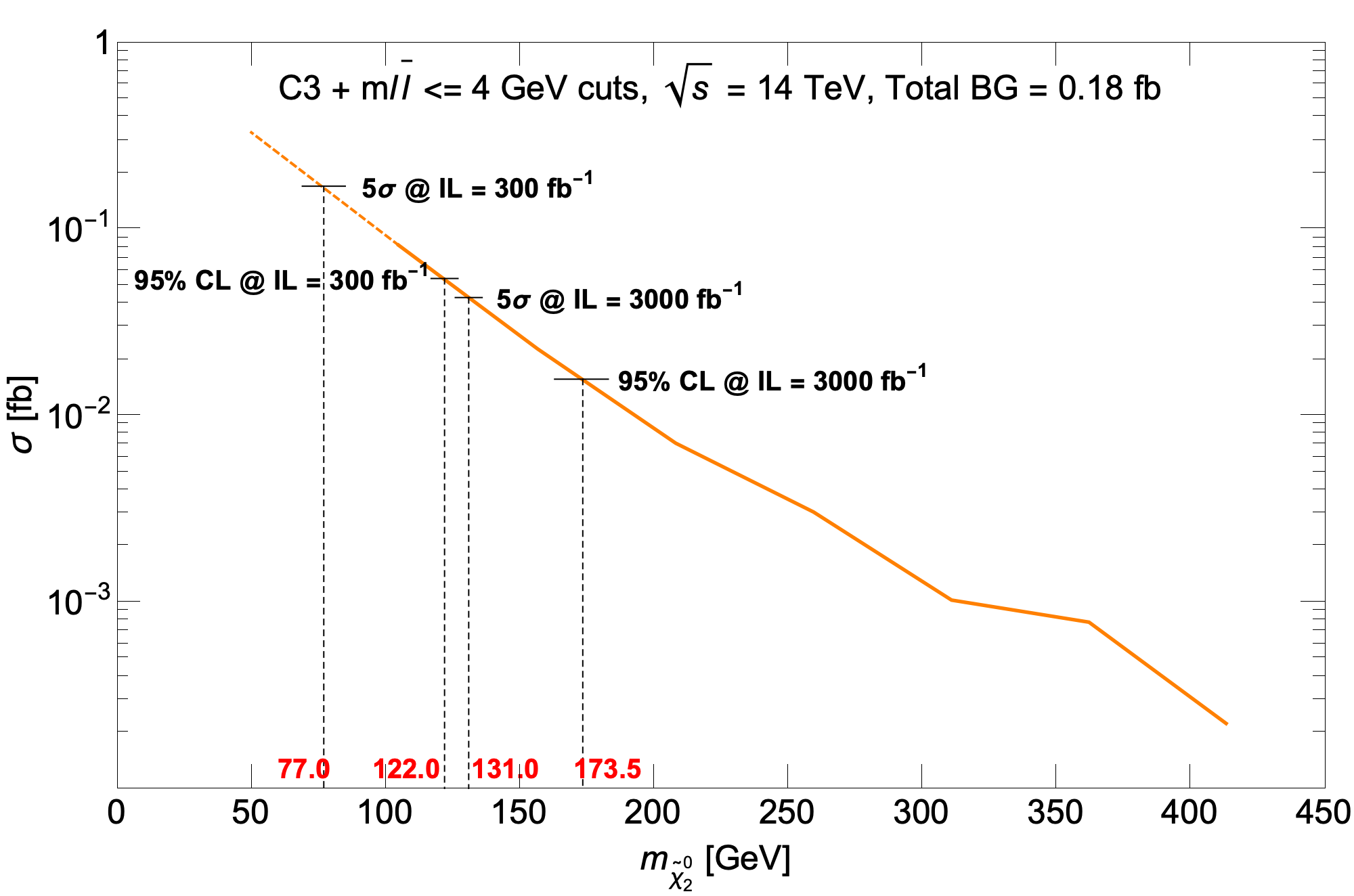}
  \caption{}
  \label{fig:mdlln_4}
\end{subfigure}%
\quad \quad
\begin{subfigure}[t]{0.44\textwidth}
  \centering
  \includegraphics[width=1.1\linewidth]{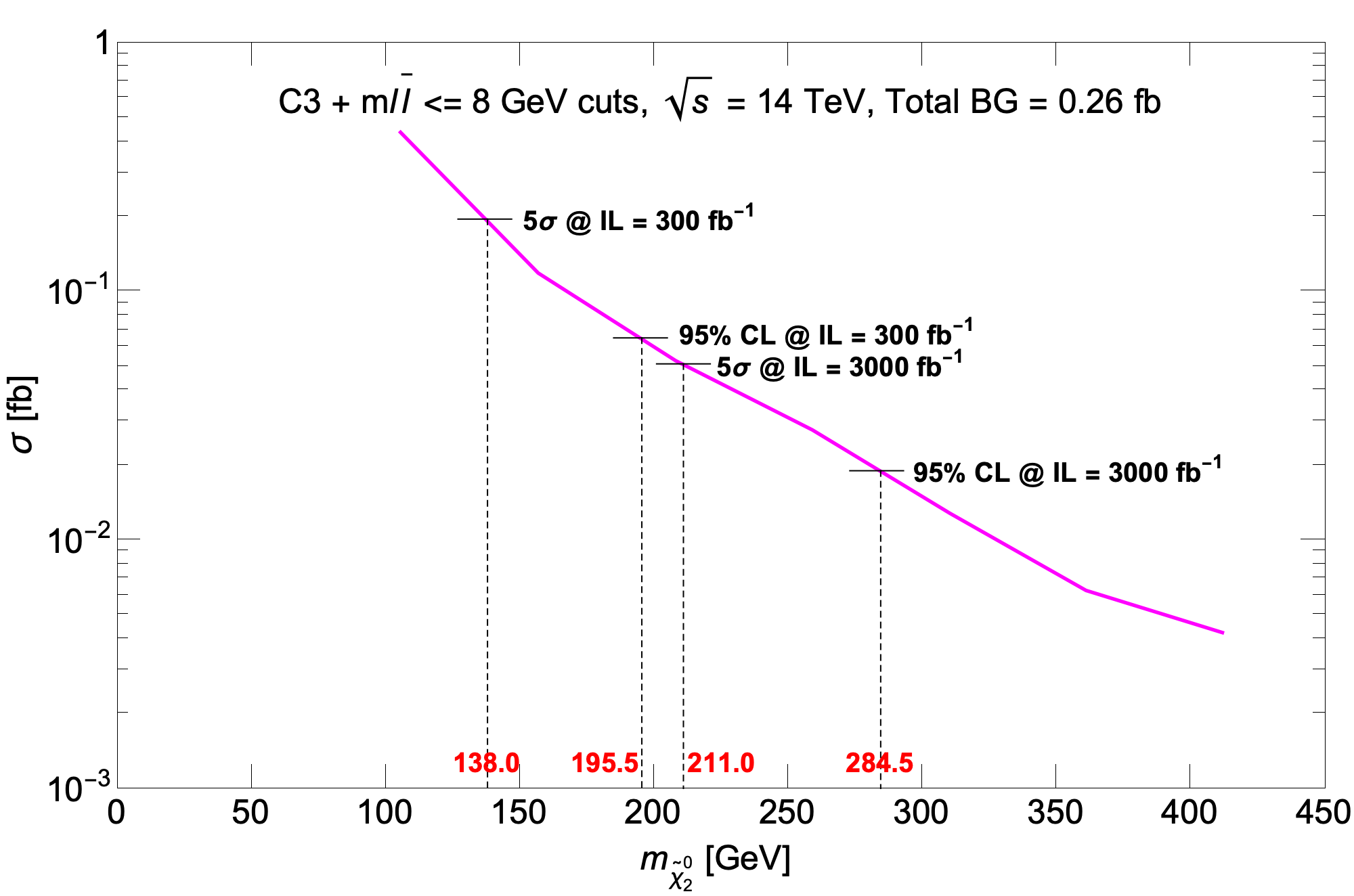}
  \caption{}
  \label{fig:mdlln_8}
\end{subfigure}
\begin{subfigure}[t]{0.44\textwidth}
  \centering
  \includegraphics[width=1.1\linewidth]{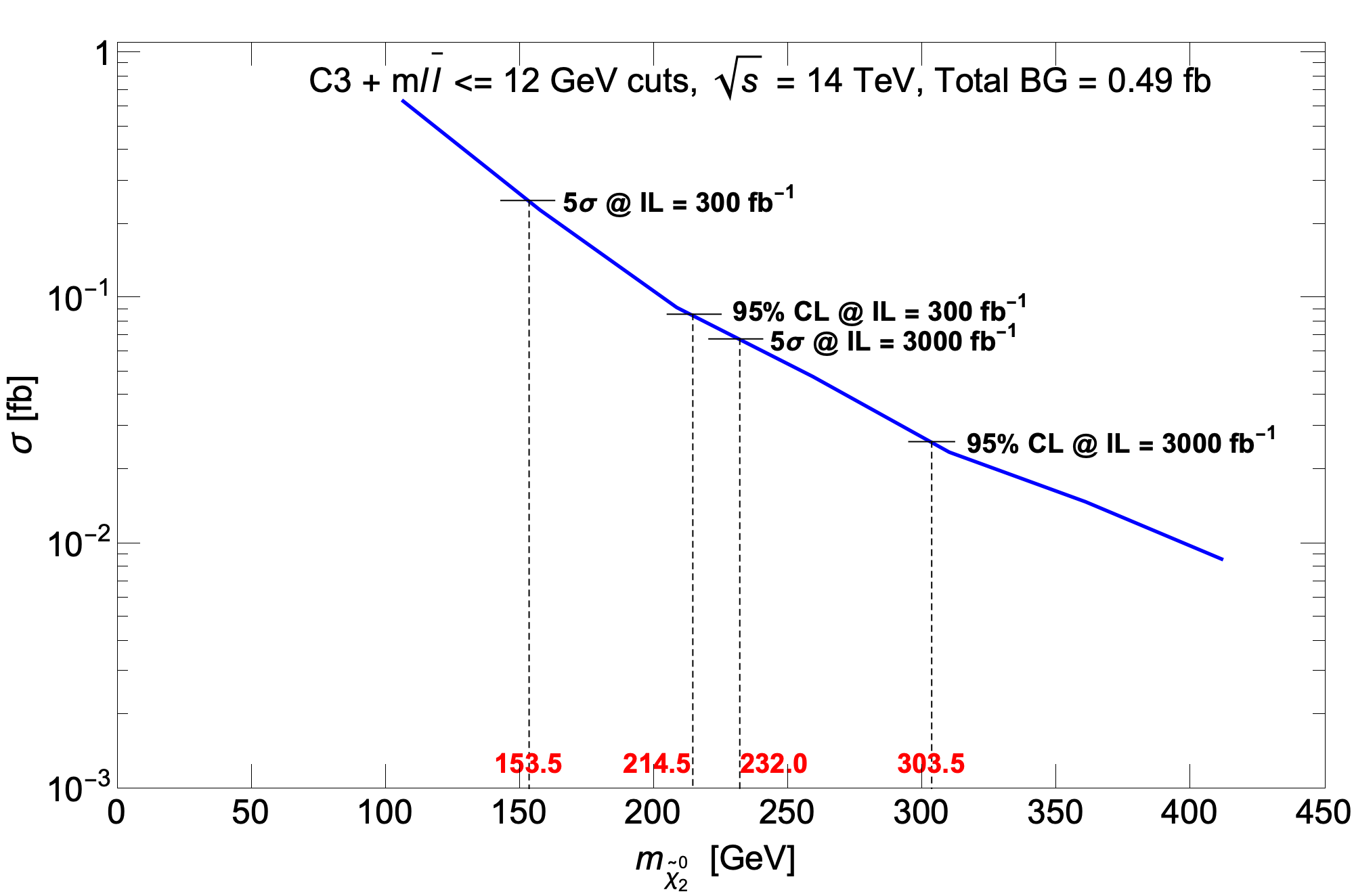}
  \caption{}
  \label{fig:mdlln_12}
\end{subfigure}%
\quad \quad
\begin{subfigure}[t]{0.44\textwidth}
  \centering
  \includegraphics[width=1.1\linewidth]{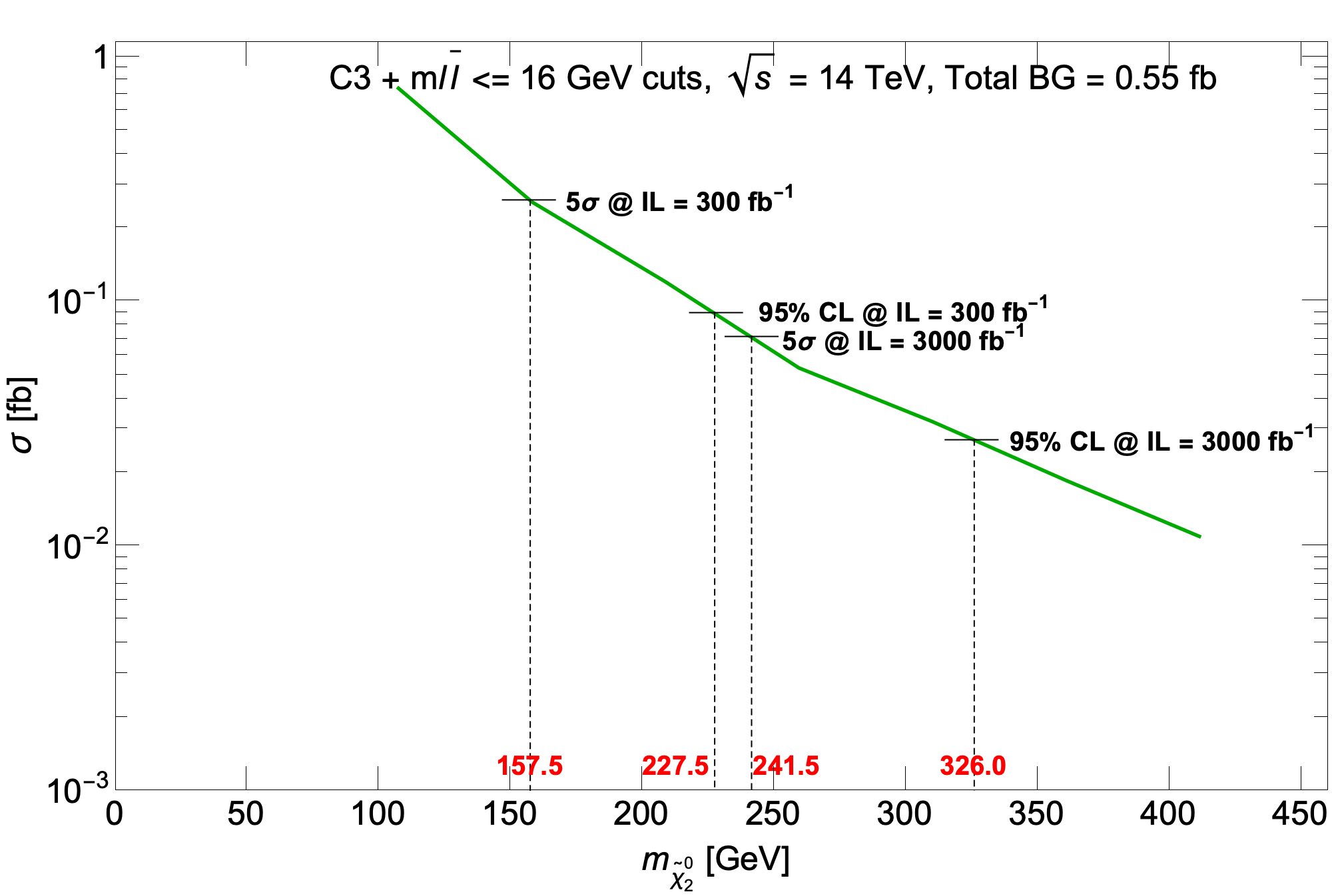}
  \caption{}
  \label{fig:mdlln_16}
\end{subfigure}
\caption{The projected $5\sigma$ reach and 95\% CL exclusion of the HL-LHC
  with 3000 fb$^{-1}$ in $\mu$ for four different NUHM2 model lines with
  {\it a}) $\Delta m=4$ GeV, {\it b}) $\Delta m=8$ GeV, {\it c}) $\Delta
  m=12$ GeV and {\it d}) $\Delta m=16$ GeV after
  $C3 + m(\ell\bar{\ell})<m_{\tz_2}-m_{\tz_1}$ cuts.
  We also list the total
  background in each frame in case the reader wishes to estimate the statistical
  significance of the signal for different choices of integrated luminosity.
\label{fig:reachC4}}
\end{figure}

Using these {\bf C4} cuts, then we computed the remaining signal cross
section after cuts for four model lines in the NUHM2 model for variable
values of $\mu :100-400$ GeV and with variable $m_{1/2}$ values adjusted
such that the $m_{\tz_2}-m_{\tz_1}$ mass gap is fixed at 4, 8, 12 and 16
GeV. While $\mu$ and $m_{1/2}$ are variable, the values of $m_0=5$ TeV,
$A_0=-1.6 m_0$, $\tan\beta =10$ and $m_A=2$ TeV are fixed for all four
model lines.\footnote{In order to get a mass gap significantly smaller
  than 10~GeV, one has to choose large $m_{1/2}$ values for which
  $\Delta_{\rm EW} > 30$. However, this is unimportant since our goal
  here is just to illustrate the reach for small mass gaps because, as
  already noted, there are top-down models with $\Delta_{\rm EW}<
  30$ and a mass gap as small as $\sim 4$~GeV. Since the signal that we
  are examining is largely determined by the lighter higgsino masses,
  the NUHM2 model serves as an effective phenomenological surrogate for our
  purpose.}  In Fig. \ref{fig:reachC4}, we show the signal cross section
after {\bf C4} cuts, along with the $5\sigma$ reach and the 95\% CL
exclusion for LHC14 with 300 and 3000 fb$^{-1}$.
We also list the total
background in each frame in case the reader wishes to estimate the statistical
significance of the signal for a given value of $m_{\tchi_2^0}$
for different choices of integrated luminosity.

In Fig. \ref{fig:reachC4}{\it a}), we find for $\Delta m=4$ GeV that the
$5\sigma$ (95\% CL) reach of LHC14 with 300 fb$^{-1}$ extends out to 80
GeV (122 GeV) respectively. For HL-LHC with 3000 fb$^{-1}$, then we
obtain the corresponding values to be 131~GeV (173.5~GeV).  Thus, the
HL-LHC should give us an extra reach in $\mu$ by $\sim 50$~GeV over the
300 fb$^{-1}$ expected from LHC Run 3.  For larger mass gaps, {\it e.g.}
$\Delta m=16$ GeV as shown in Fig. \ref{fig:reachC4}{\it d}), then the
signal is larger, but so is background since now we require a larger
$m(\ell\bar{\ell})$ signal bin.  For $\Delta m=16$~GeV, the 300
fb$^{-1}$ reach is to 157.5~GeV (227.5~GeV) respectively.  For 3000
fb$^{-1}$, the corresponding reach (exclusion)  extends to 241.5~GeV 
(325~GeV).  Thus, the reach is largest for the larger mass gaps, as might be
expected.  The intermediate mass gaps give LHC mass reaches in between
the values obtained for the lower and higher $\Delta m$ values.

\begin{figure}[!htbp]
\begin{center}
\includegraphics[height=0.4\textheight]{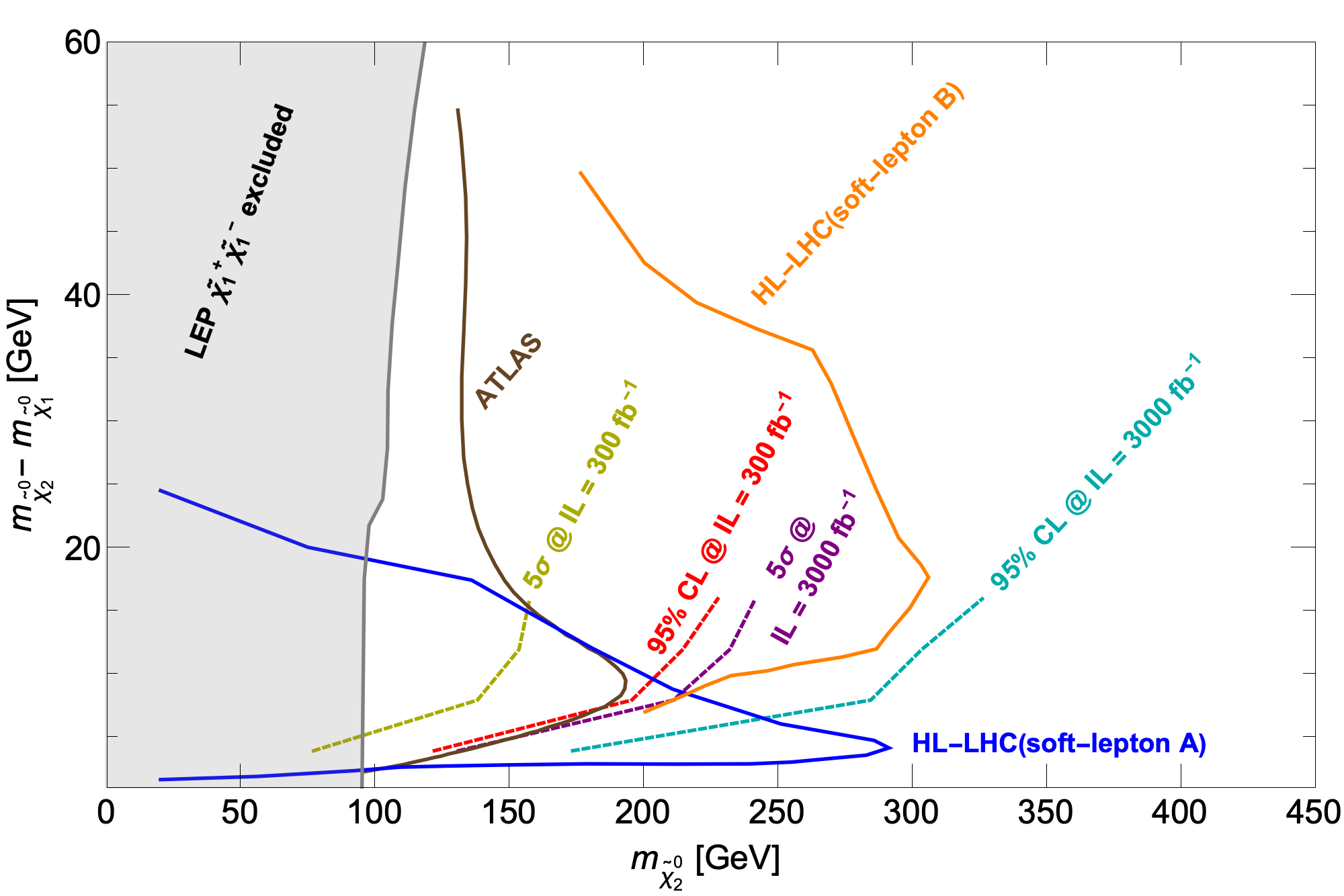}
\caption{The projected $5\sigma$ reach and 95\% CL exclusion contours
  for LHC14 with 300 and 3000 fb$^{-1}$ in the $m_{\tz_2}$ vs. $\Delta
  m$ plane after $C4$ cuts. Also shown is the current 95\% CL exclusion
  (ATLAS) and the projected 95\% CL exclusions from two different
  analyses for the HL-LHC \cite{Canepa:2020ntc}.
\label{fig:reach}}
\end{center}
\end{figure}

In Fig. \ref{fig:reach}, we translate the results of Fig.~\ref{fig:reachC4}
into the standard $m_{\tz_2}$ vs. $\Delta m$ plane. We also show the region 
excluded by LEP2 chargino searches (gray region).
Also shown is current 95\%CL exclusion region (labelled ATLAS) along
with the projections of what searches at the HL-LHC would probe at the
95\%CL \cite{Canepa:2020ntc}: ATLA (soft-lepton A) and CMS (soft-lepton B). 
We see that the reach that we obtain
compares well with the corresponding projections by the ATLAS and CMS
collaborations. Our focus here has been on higgsino mass gaps $\alt
20-25$~GeV, expected in natural SUSY models. For larger mass gaps, the
search strategy explored in this paper becomes less effective because of
increased backgrounds from $t\bar{t}$, $WWj$ and other SM processes, and
the reach contours begin to turn over. In this case, it may be best to
search for higgsinos via the hard multilepton events, without the need
for a QCD jet.

Before closing this section, we note that we have only considered
physics backgrounds in our analysis. The ATLAS collaboration has,
however, reported that a significant portion of the background comes
from fake leptons, both $e$ and $\mu$. Accounting for these
detector-dependent backgrounds (which may well be sensitive to the
HL-LHC environment as well as upgrades to the detectors) require data
driven methods which are beyond the scope of our study. We point out,
however, that the reader can roughly gauge the impact of the fakes on
the contours shown in Fig.~\ref{fig:reach} using the curves in
Fig.~\ref{fig:reachC4}. For instance, if the fakes increase the
background by a factor $f$, the cross section necessary to maintain the
same significance for the signal would have to increase by $\sqrt{f}$;
i.e. if the fakes doubled the background, for $\Delta m$=8~GeV, the HL-LHC
discovery limit would reduce by $\sim 25$~GeV. 
In the same vein, the reach would be increased by $\sim 30$~GeV
if the data from the two experiments could be combined.

\section{Conclusions}
\label{sec:conclude}

It is generally agreed that naturalness in supersymmetric models
requires the SUSY preserving higgsino mass $\mu$ rather nearby to the
weak scale, because it enters Eq.~(\ref{eq:mzsq}) at tree level. The
soft SUSY breaking parameters, however, may be well beyond the TeV scale without
compromising naturalness as long as $m_{H_u}^2$ is driven to small
negative values at the weak scale.  Indeed, a subset of us
\cite{land0,land1,land2,land3} have advocated that anthropic
considerations on the string landscape favour large values of soft SUSY
breaking parameters, but not so large that their contributions to the
weak scale are too big.  Such a scenario favours $m_h\sim 125$ GeV with
sparticles other than higgsinos well beyond HL-LHC reach. While {\it
  stringy naturalness} provides strong motivation for higgsino pair
production reactions as the most promising avenue to SUSY discovery at
LHC14, the phenomenological analysis presented in this paper applies to
any MSSM framework with a compressed spectrum of light higgsinos. 

We have re-examined the prospects for a search for soft
opposite-sign/same flavor dilepton plus $\eslt$ from higgsino pair
production in association with a hard monojet at LHC with $\sqrt{s}=14$ TeV.
The dileptons originate from $\tz_2\to\ell\bar{\ell}\tz_1$ so that the
dilepton pair has a distinctive kinematic edge with
$m(\ell\bar{\ell})<m_{\tz_2}-m_{\tz_1}$, while the monojet serves as the
event trigger.

We examined several signal benchmark cases, and compared the signal
against SM backgrounds from $t\bar{t}$, $\tau\bar{\tau}j$, $WWj$,
$W\ell\bar{\ell}j$ and $Z\ell\bar{\ell}j$ production.  The ditau mass
reconstruction $m_{\tau\tau}^2$, valid in the collinear tau decay
approximation for decays of relativistic taus, has been used to reduce
the dominant background from $Z(\to\tau\bar{\tau})+j$ production.
However, significant ditau background remains even after the
$m_{\tau\tau}^2<0$ cut. In this
paper, we proposed a new set of angular cuts which eliminate ditau
backgrounds much more efficiently at relatively low cost to
signal. Additional analysis cuts allow for substantial rejection of
$t\bar{t}$ and other SM backgrounds. In the end, we expect higgsino pair
production to manifest itself as a low end excess in the
$m(\ell\bar{\ell})$ mass distribution with a cutoff at the $\Delta
m=m_{\tz_2}-m_{\tz_1}$ value, with a tail extending to larger values of
$m(\ell\bar{\ell})$ when the two leptons originate in different
higgsinos. Using the so-called {\bf C3}$+m(\ell\bar{\ell})$ cuts, we
evaluated the reach of LHC14 for 300 and 3000 fb$^{-1}$ of integrated
luminosity.

Our final result is shown in Fig.~\ref{fig:reach}. We see that the reach
is strongest for larger $\Delta m$ values up to $15-20$ GeV but drops
off for smaller mass gaps. 
Mass gaps smaller than about 4 GeV occur only for very heavy gauginos 
that fail to satisfy our naturalness criterion, 
while higgsinos with an uncompressed spectrum
would have large mixing with the electroweak gauginos and can be more
effectively searched for via other channels.  We see from
Fig.~\ref{fig:reach} that the HL-LHC with 3000 fb$^{-1}$ gives a
$5\sigma$ discovery reach to $m_{\tz_2}\sim 240$ GeV, with the 95\% CL
exclusion limit extending to $\sim 325$ GeV for $\Delta m\sim 16$
GeV. Nonetheless, a significant portion of natural parameter space with
$\mu\sim m_{\tz_2}\sim 200-350$ GeV and $\Delta m\sim 4-10$ GeV may
still be able to evade HL-LHC detection.  Given the importance of this
search, we urge our experimental colleagues to see if it is possible to
reliably extend the lepton acceptance to yet lower $p_T$ values, or
increase $b$-quark rejection even beyond 80-85\% that has already been
achieved.

{\it Acknowledgements:} 

This work has been performed as part of a contribution to the Snowmass 2022
workshop.
This material is based upon work supported by the U.S. Department of Energy, 
Office of Science, Office of Basic Energy Sciences Energy Frontier Research 
Centers program under Award Number DE-SC-0009956 and U.S. Department of Energy 
Grant DE-SC-0017647. The work of DS was supported by the Ministry of Science 
and Technology (MOST) of Taiwan under Grant No. 110-2811-M-002-574.


%
\end{document}